\documentclass[final, 5p, sort&compress, times, twocolumn]{elsarticle}
\usepackage{graphicx}
\usepackage{wrapfig, framed, caption}
\usepackage[cmex10]{amsmath}
\usepackage{stackengine}

\usepackage{float}
\usepackage{amssymb}
\usepackage{siunitx}
\usepackage{cleveref}
\allowdisplaybreaks
\usepackage[toc,nopostdot,nonumberlist]{glossaries}
  
\usepackage{acro}
\makeglossaries
\usepackage[latin1]{inputenc}
\usepackage{tikz}
\usetikzlibrary{shapes,arrows}
\usepackage[normalem]{ulem}
\useunder{\uline}{\ul}{}
\usepackage{algorithm,algcompatible,amsmath}
\setcitestyle{square}
\setcitestyle{comma}
\journal{Applied Energy}

\begin{document}
\begin{frontmatter}
\title{A Novel Probabilistic Framework to Study the Impact of PV-battery Systems on \\ Low-Voltage Distribution Networks}

\author[label1]{Yiju Ma\corref{cor1}}
\address[label1]{School of Electrical and Information Engineering, The University of Sydney, Sydney, Australia}

\cortext[cor1]{Corresponding author}


\ead{yiju.ma@sydney.edu.au}

\author[label1]{Donald Azuatalam}
\ead{donald.azuatalam@sydney.edu.au} 
 
\author[label1]{Thomas Power}
\ead{thomas.power@sydney.edu.au}

\author[label1]{Gregor Verbi\v{c}}
\ead{gregor.verbic@sydney.edu.au}

\author[label1]{Archie~C. Chapman}
\ead{archie.chapman@sydney.edu.au}

\begin{abstract}
Battery storage, particularly residential battery storage coupled with rooftop PV, is emerging as an essential component of the smart grid technology mix. 
However, including battery storage and other flexible resources like electric vehicles and loads with thermal inertia into a probabilistic analysis based on Monte Carlo (MC) simulation is challenging, because their operational profiles are determined by computationally intensive optimization. Additionally, MC analysis requires a large pool of statistically-representative demand profiles to sample from. As a result, the analysis of the network impact of PV-battery systems has attracted little attention in the existing literature.
To fill these knowledge gaps, this paper proposes a novel probabilistic framework to study the impact of PV-battery systems on low-voltage distribution networks. Specifically, the framework incorporates home energy management (HEM) operational decisions within the MC time series power flow analysis. First, using available smart meter data, we use a Bayesian nonparametric model to generate statistically-representative synthetic demand and PV profiles. 
Second, a policy function approximation that emulates battery scheduling decisions is used to make the simulation of optimization-based HEM feasible within the MC framework. The efficacy of our method is demonstrated on three representative low-voltage feeders, where the computation time to execute our MC framework is 5\% of that when using explicit optimization methods in each MC sample. 
The assessment results show that uncoordinated battery scheduling has a limited beneficial impact, which is against the conjecture that batteries will serendipitously mitigate the technical problems induced by PV generation.
\end{abstract}


\end{frontmatter}


\section{Introduction} \label{intro}
Residential rooftop PV constitutes an increasingly important part of the electricity supply mix. In Australia, for example, the annual installed capacity of small-scale PV systems has grown from less than \SI{200}{MW} in 2009 to \SI{1.1}{GW} in 2017, with the average installation size rising from \SI{1.5}{kW} to \SI{5.5}{kW} \cite{aemosmall}.
Improvements in small-scale battery storage technologies continue apace, and battery storage is becoming a popular technology in countries with a high penetration of rooftop PV. 
In Australia, \SI{12} {\%} of the {172,000} residential solar installations in 2017 included a battery, while this proportion was only \SI{5}{\%} in 2016. Given this, a total of {28,000} battery systems had been installed by the end of 2017 \cite{climatefc}. In addition, projections by AEMO see residential battery capacity reaching \SI{6.6}{GW} by 2035, with \SI{3.8}{GW} expected to be installed as part of PV-battery systems \cite{AEMO4}. A key driver to this trend is their falling costs, which are predicted to drop by \SI{50}{\%} from 2017 to 2037 \cite{AEMO3}. 
A similar rapid storage deployment has been observed in Germany. According to the German Solar Industry Association, more than 100,000 solar battery system were installed in 2018, and more than a half new PV installations come with battery storage.

Rooftop PV penetration levels have in many jurisdictions reached levels where network issues have started to emerge; these include over-voltages, reverse power flows, and phase unbalance \cite{tonkoski2012impact}. At the same time, it is widely conjectured that battery systems will largely mitigate the problems of excessive PV generation \cite{BrinsmeadEtAl-ENA-CSIRO-Transformation-Roadmap2017}. However, battery storage and other flexible resources like electric vehicles and loads with thermal inertia pose a challenge in power system planning in operation. This is because their operational profiles are determined by the tools of optimization. To properly capture the varying system conditions requires optimization horizons of up to a day or even more, which is computationally burdensome. For this reason, including flexible resources in the existing Monte Carlo (MC) approaches, commonly used in probabilistic power system analysis, becomes infeasible.

Very little work has been done so far to address the issue of how to include optimization in MC simulation. A possible solution is to reformulate the problem, for example by using a two-point estimate method as in \cite{verbic2006probabilistic}, but this requires some strong assumptions, such as modeling random variables by well-behaved probability distributions. Using such assumptions to properly capture the stochastic behavior of residential users with PV-battery systems is overly simplistic; in fact, it requires the use of more sophisticated modelling techniques, such as Bayesian nonparametric models. 
However, no existing research has investigated the extent to which hybrid PV-battery systems can benefit distribution networks with high PV penetration, due to the infeasibility of the resulting MC simulation, as discussed above. Because grid integration of large populations of behind-the-meter distributed energy resources has become a hot topic in Australia \cite{OpenEnergyNetworks}, methods and tools to study the impact of PV-battery systems on low-voltage (LV) distribution networks are desperately needed.

This paper fills this important knowledge gap by proposing a novel probabilistic framework that: (1) overcomes the drawback of existing MC approaches that cannot explicitly include optimization due to its excessive computational burden; and (2) provides a principled statistical way to generate a large number of residential demand and PV traces required for MC analysis. 
The proposed framework has a broad application appeal in the power system context; it can be used in applications that require optimization for resource scheduling, such as probabilistic studies of the impact of distributed energy resources (DER) scheduling on distribution networks (the focus of this paper), economic appraisal of DER investments in network planning using real options analysis, and probabilistic estimation of DER capacity available for power system frequency and voltage services.

\subsection{Related Work} \label{litrev}
In this subsection we review the existing methods for: (i) assessing the impact of PV generation on distribution networks using MC analysis; (ii) solving the battery scheduling problem with a home energy management (HEM) system; and (iii) modeling customers' solar generation and electrical demand.

\subsubsection{Monte Carlo Approaches}
MC analysis has been employed in many studies to capture uncertainties in the location and size of DER when assessing their impacts on voltage profiles and peak loading in LV networks \cite{chen2012analysis, vallee2013development, navarro2014probabilistic,navarro2016probabilistic,protopapadaki2017heat}. In particular, \cite{chen2012analysis} applied a probabilistic approach to evaluate the impacts of distributed generation with different penetration levels on voltage profiles. The authors in \cite{navarro2014probabilistic} proposed a probabilistic methodology based on MC analysis to investigate the impact of electric heat pumps on LV networks, with a focus on voltage and thermal limits. A similar approach was used in \cite{navarro2016probabilistic} to probabilistically allocate PV, combined heat and power systems, electric heat pumps and electric vehicles (EVs), to investigate the prevalence of voltage problems on LV feeders with different penetration levels of these \textit{low carbon technologies}. The authors in \cite{protopapadaki2017heat} studied the impact of PV generation and electric heat pumps on LV networks using MC analysis that captures uncertainties in building characteristics (geometry and insulation quality), feeder size, cable type and heat pump and PV penetration levels. In addition, \cite{kolenc2015assessment} proposed a probabilistic approach using MC analysis to investigate the maximum PV penetration level that can be tolerated by an LV network without voltage problems. However, none of these studies includes battery storage. Furthermore, they assume that schedules of flexible devices are independent of PV generation, which might not be true in practice. Charging of EVs, for example, should preferably coincide with co-located PV generation to minimize grid in-feed. Additionally, flexible technologies with thermal inertia can serve as ''solar sponges'' to maximize self-consumption \cite{rahimpour2017using}.  


\subsubsection{Home Energy Management}
Battery scheduling determines the ``behavior'' of the PV-battery system and hence the impact on the network. The scheduling problem is typically formulated as an optimization problem and solved using a HEM system. The objective can vary but it is usually to minimize energy expenditure for the user \cite{Barbato2014}. In jurisdictions where the PV buyback tariff is less than the electricity tariff (e.g. in Australia), energy expenditure minimization coincides with self-consumption maximization.

Several methods have been proposed to solve the HEM problem, including stochastic mixed integer linear programming \cite{chen2012real,erdinc2015new}, mixed integer quadratic programming \cite{killian2018comprehensive}, dynamic programming (DP) \cite{tischer2011towards} and approximate dynamic programming \cite{keerthisinghe2016fast}. A common feature of these methods is a relatively high computational burden, which hinders a direct implementation in a MC framework.
To improve the computational performance, \cite{Chanaka2018} proposed a \textit{policy function approximation} (PFA) algorithm, which uses the battery schedules from the solution of the HEM problem to train an artificial neural network (ANN) that maps demand and PV generation to battery scheduling output. The ANN is then used as a PFA to obtain the battery schedule without having to solve the underlying optimization problem. 

\subsubsection{Solar and Demand Modeling}
For synthesizing stochastic demand and PV profiles, Markov chains using a bottom-up approach starting at the appliance level are typically used. Examples include the simulation of building occupancy profiles for the purposes of generating lighting demand \cite{widen2009combined} and residential energy demand profiles \cite{richardson2010domestic,widen2010high}. However, such bottom-up approaches are computationally expensive, and it is difficult to model energy usage at the appliance level in a way that represents the diversity of customer behavior. 

Given these shortcomings, \cite{2018arXiv180800615P} proposed a methodology for generating residential demand and solar profiles using a Markov process specific to the features of the existing \textit{smart meter} data. 
Instead of working up from the appliance level, the method generates the synthetic profiles by clustering a set of observed profiles using a Dirichlet process, and then generating transition matrices used in the Markov process from these clusters. 

\subsection{Contributions}
Within this context, this work proposes a novel probabilistic impact assessment framework that embeds the battery scheduling optimization problem in the MC analysis to assess the effects of battery scheduling on LV networks. The incorporation of battery scheduling in such analysis has not been addressed in the existing literature, for two reasons: (i) a large number of residential demand and PV traces is needed for the MC analysis, and (ii) incorporating the HEM problem within the MC analysis, which is impractical due to the excessive computational burden. 
To solve these problems, we build on our previous work on PFA \cite{Chanaka2018} and Bayesian modeling \cite{2018arXiv180800615P} by combining them into a unified MC framework
\footnote{Due to the space constraints we couldn't explain our previous work \cite{Chanaka2018, 2018arXiv180800615P} in full detail. Nevertheless, the summary provided below is detailed enough to make the paper sufficiently self-contained.}. Specifically:
\begin{itemize}
  \item We use a Markov chain approach to synthesize large numbers of statistically similar, but independent demand and PV profiles using smart meter data;
  \item We incorporate battery scheduling optimization within the MC analysis by training a PFA using an ANN to estimate the near-optimal battery schedules for a large pool of customers, which reduces the computational time required to solve the HEM problem by more than \SI{95}{\%};
  \item We complete a comprehensive PV hosting capacity assessment using a probabilistic time-series power flow analysis for several representative LV feeders to show that uncoordinated battery scheduling has a limited beneficial impact on LV networks; this disproves the conjecture that battery storage will serendipitously mitigate the technical problems induced by PV generation. 
\end{itemize}

The proposed framework overcomes the shortcoming of existing MC power flow studies, which fail to accommodate schedulable batteries. Thus, for the first time, our work provides a MC framework that explicitly includes the DER scheduling used to manage customers' behind-the-meter energy use. 
Furthermore, the proposed framework allows MC power flow analysis to be conducted with exiguous smart meter data. Smart-meter roll-out is ongoing, but there are still many jurisdictions where coverage is still patchy, e.g. NSW in Australia \cite{CIGRE}. 
By filling these existing gaps, the proposed framework provides a tool-chain for technically and financially evaluating future DER deployment and investment, as we demonstrate in a rooftop PV hosting capacity assessment.
Moreover, our framework can be used to develop other types of future network assessments and investigations, including real options valuations of staged DER deployment and probabilistic DER power system service capacity estimation.  


The remainder of the paper is organized as follows. Section \ref{section2} describes the module that synthesizes large pools of demand and PV profiles. Section \ref{section3} presents the HEM system used to compute the battery schedules for each customer from the large data pool. The probabilistic power flow study via MC analysis is described in Section \ref{section4}. The three modules described in Sections \ref{section2}, \ref{section3} and \ref{section4} are used to form the probabilistic impact assessment framework, shown in Fig.~\ref{fig2}. This framework is used to assess the impacts of battery scheduling under \textit{time-of-use} tariffs and \textit{self consumption maximization} on mitigating network issues, and the results are discussed in Section \ref{section5}. Section \ref{section6} draws conclusions.

\pgfdeclarelayer{background}
\pgfdeclarelayer{foreground}
\pgfsetlayers{background,main,foreground}

\tikzstyle{materia}=[draw, thick, text width=30em, text centered,
minimum height=1.5em]
\tikzstyle{practica} = [materia, fill=blue!20, text width=28em, minimum width=28em,
minimum height=3em, rounded corners]
\tikzstyle{texto} = [above, text width=20em]
\tikzstyle{linepart} = [draw, thick, -latex', dashed]
\tikzstyle{line} = [draw, thick, -latex']
\tikzstyle{ur}=[draw, text centered, minimum height=0.01em]

\newcommand{\blockdist}{1.3}
\newcommand{\edgedist}{1.5}

\newcommand{\practica}[2]{node (p#1) [practica]
	{\textbf{Step #1}\\{\normalsize\textit{#2}}}}

\newcommand{\background}[5]{%
	\begin{pgfonlayer}{background}
		\path (#1.west |- #2.north)+(-0.6,0.6) node (a1) {};
		\path (#3.east |- #4.south)+(+0.6,-0.6) node (a2) {};
		\path[fill=yellow!20, rounded corners, draw, thick, dashed]
		(a1) rectangle (a2);
		\path (a1.east |- a1.south)+(4,-0.4) node (u1)[texto]
		{\normalsize\textit{Module #5}};
\end{pgfonlayer}}

\newcommand{\transreceptor}[3]{%
	\path [linepart] (#1.east) -- node [above]
	{\scriptsize Transreceptor #2} (#3);}
    
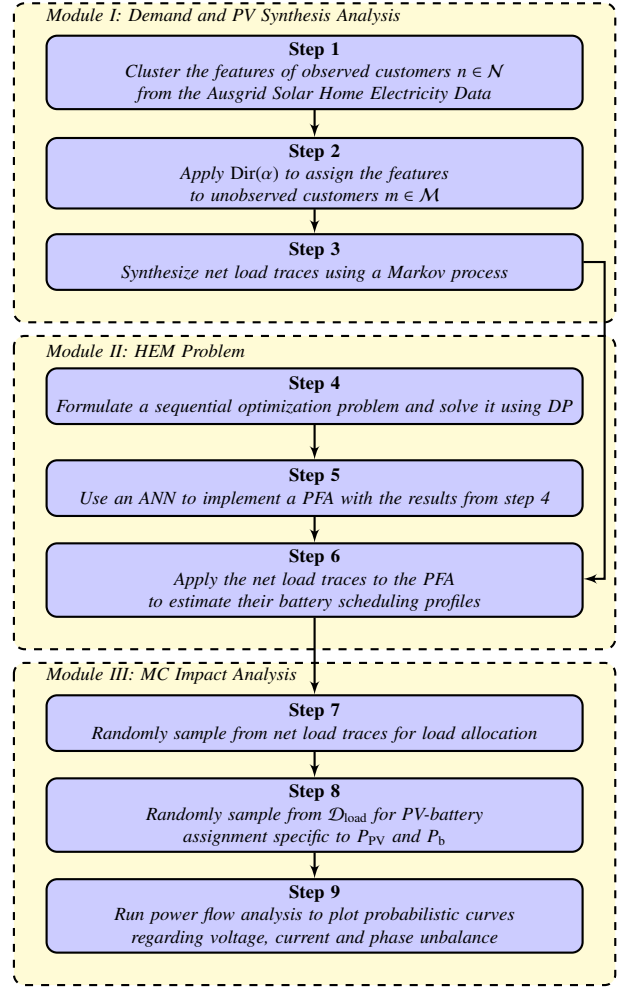
\begin{figure}[t]
	\begin{center}

	\begin{tikzpicture}[scale=0.7,transform shape]
	
	\path \practica {1}{Cluster the features of observed customers ${n \in \mathcal{N}}$ from the Ausgrid Solar Home Electricity Data};
	\path (p1.south)+(0.0,-1.2) \practica{2}{Apply $\operatorname{Dir}(\alpha)$ to assign the features \\ to unobserved customers ${m \in \mathcal{M}}$};
	\path (p2.south)+(0.0,-1.0) \practica{3}{Synthesize net load traces using a Markov process};

	\path (p3.south)+(0.0,-2.0) \practica{4}{Formulate a sequential optimization problem and solve it using DP};
	\path (p4.south)+(0.0,-1.2) \practica{5}{Use an ANN to implement a PFA with the results from step 4};
	\path (p5.south)+(0.0,-1.2) \practica{6}{Apply the net load traces to the PFA \\ to estimate their battery scheduling profiles};
	
	\path (p6.south)+(0.0,-2) \practica{7}{Randomly sample from net load traces for load allocation};
	\path (p7.south)+(0.0,-1.2) \practica{8}{Randomly sample from ${\mathcal{D}_\mathrm{load}}$ for PV-battery \\ assignment specific to $P_\mathrm{PV}$ and $P_\mathrm{b}$};
	\path (p8.south)+(0.0,-1.2) \practica{9}{Run power flow analysis to plot probabilistic curves regarding voltage, current and phase unbalance};

	\path [line] (p1.south) -- node [above] {} (p2);
	\path [line] (p2.south) -- node [above] {} (p3);

     \path [line] (p3.east) -- +(0.4,0.0) -- +(0.4,-6) -- node [right] {} (p6);
	\path [line] (p4.south) -- node [above] {} (p5);

	\path [line] (p5.south) -- node [above] {} (p6);
	\path [line] (p6.south) -- node [above] {} (p7);
	
	\path [line] (p7.south) -- node [above] {} (p8);
	\path [line] (p8.south) -- node [above] {} (p9);

	\background{p1}{p1}{p1}{p3}{I: Demand and PV Synthesis Analysis}
	\background{p4}{p4}{p4}{p6}{II: HEM Problem}
	\background{p1}{p7}{p8}{p9}{III: MC Impact Analysis}

	\end{tikzpicture}
    	\caption{Overview of the Methodology.}
	    \label{fig2}
\end{center}
\end{figure}
 
\section{Demand and PV Trace Models} \label{section2}
Probabilistic assessment of the impact of DER on LV distribution networks requires a large pool of synthetic demand and PV traces to sample from. In many jurisdictions smart metering data are scarce, so we need to be able to generate statistically representative samples also when only a limited number of demand and PV traces is available \cite{CIGRE}.
This is the purpose of Module 1 in Fig.~\ref{fig2}; to use an existing data set to generate a larger pool of demand and PV profiles (net load traces). Module~1 works by assigning Markov processes according to a Dirichlet distribution identified via clustering, as explained below.

\subsection{Data Preparation}
This work extends the non-parametric Bayesian model \cite{2018arXiv180800615P} to generate net load traces that are statistically similar to historical demand and PV generation of observed customers. The observed data was collected from the \textit{Ausgrid Solar Home Electricity Data} \cite{SHED}. 

Let $n \in \mathcal{N}$ and $m \in \mathcal{M}$ denote the set of observed and unobserved customers, respectively. The module first applies a clustering technique, namely \textit{maximum a-posteriori Dirichlet process mixtures} \cite{raykov2016k}, to cluster $n \in \mathcal{N}$ customers into representative sets $k \in \mathcal{K}$ according to their features. This technique is useful for instances in which the number of clusters cannot be easily determined. The features of demand are the day types (weekday or weekend) and number of residents, while those for PV include the PV capacity, panel orientation and weather information. Clustering is important because (i) considering each customer as a single category is computationally expensive, and (ii) it provides generalizable statistical information as the demand and PV generation in each set are correlated with their features. 

\subsection{Estimating the Dirichlet Distribution}
After clustering, we could compute the frequencies, $\{p_k\}_{k \in \mathcal{K}}$,  of each $k \in \mathcal{K}$ in the population $\mathcal{N}$. These frequencies can be interpreted as the probability of an unobserved customer having certain features. However, they are only an estimate across the observed customers, and directly using them to allocate features fails to properly consider the error in this estimate, which can be significant where the fraction of customers observed is small. Thus, a Bayesian estimation approach is employed.

Specifically, in Step 2 in Module 1, the model uses the count of each $k \in \mathcal{K}$ in the observed $\mathcal{N}$ as a hyperparameter of a  Dirichlet distribution, which itself is sampled to yield a  \textit{categorical} probability distribution over the features for unobserved customers, ${m \in \mathcal{M}}$.  Formally, this is given by: 

\[ {\displaystyle {\begin{array}{rcl}
\boldsymbol{\alpha }& & \text{Vector of cluster counts}\\
\mathbf {p} \mid \boldsymbol {\alpha }& \sim &\operatorname {Dir} (\boldsymbol {\alpha })\\
K_m \mid \mathbf {p} & \sim &\operatorname {Cat} (\mathbf{p} )
\end{array}}}
\]

In more detail, $\boldsymbol {\alpha}$ is a vector of \textit{concentration hyper-parameters} given by the number (c.f.~frequency) of observed customers within each $k \in \mathcal{K}$. Sampling from $\operatorname{Dir}(\boldsymbol{\alpha})$ yields the parameters, $\mathbf{p}$ of a {categorical} probability distribution, $\operatorname{Cat}(\mathbf{p})$ over the features for unobserved customers, ${m \in \mathcal{M}}$. Finally, $K_m$ is the random variable assigning a cluster to each unobserved customer $m$, which is drawn from $\operatorname{Cat}(\mathbf{p})$. This Bayesian approach to assigning clusters to unobserved customers ensures that the error in the estimate previously discussed is probabilistically accounted for.

\subsection{Markov Chain Process}
Step 3 in Module 1 involves synthesizing a large number of net load traces based on the feature assignments, by (i) generating Markov transition matrices and then (ii) sampling a trace, as follows. 

First, a time-inhomogeneous Markov process is identified by constructing a set of observed state transition matrices, $\{T_{n,t}\} \ \forall\, n, t$; each of which is indexed by the states for one observed customer for one time-step. Following this, a matrix of transition frequencies that records all observed state transitions is constructed for each $T_{n,t}$. An unobserved state transition matrix, denoted $T_{m,t}$, is generated specific to each of these transition frequency matrices. For this study, a time-step of 30 minutes is used as such, 48 different state transition matrices will be required to sample a profile over a one day. 

Synthetic profiles can be directly sampled from the transition matrices $T_{m,t}$, by first defining the initial state. Specifically, we sum each row of the first transition matrix, and run Gaussian kernel density estimation over these sums to give a probability measure; the initial state is then sampled from a categorical distribution defined by this measure. 
Gaussian kernel density estimation ensures that unobserved state transitions are attainable (i.e. all states communicate and the Markov chain is recurrent). Given this, we run kernel estimation over each row of each transition matrix to provide the probability measures for the rest of the states. This process is continued for each remaining time-step to construct one net load trace for one year. The net load traces serve as the inputs to the HEM problem in Module 2 for fast battery scheduling estimation. For more details regarding the non-parametric Bayesian model, please see \cite{2018arXiv180800615P}.

\section{Home Energy Management} \label{section3}
The choice of the HEM optimization formulation is arbitrary; none of the existing solution techniques is computationally efficient enough to be directly used in MC analysis. In this work (Module 2) we used DP\footnote{DP is notorious for the curse of dimensionality, but in this study we consider deterministic DP with only one schedulable device, so the computational performance is comparable to mixed-integer linear programming.} in conjunction with PFA \cite{Chanaka2018} to emulate battery scheduling policies for the large pool of customers synthesized in Module 1, allowing the HEM operational decisions to be feasibly included within the MC analysis. Specifically, the process first formulates a \textit{Markov decision process} (MDP) for each observed customer, $n\in \mathcal{N}$. The objective is to minimize the energy costs for each customer, with costs and benefits given by time-of-use tariffs and feed-in-tariffs. The decision variables of this algorithm include the optimal scheduling policies for each battery system over a year. Next, using the observed data set and the outputs from solving the HEM problem, an ANN is trained as a PFA algorithm, which is then used to compute fast solutions to the battery scheduling problem for the net load traces synthesized in Module 1. The details of Module 2 are discussed below. 

\subsection{Scheduling Problem}

The general formulation of the scheduling problem for each HEM system follows \cite{keerthisinghe2016fast}. In brief, the battery scheduling problem comprises a sequence of time-steps, $\mathcal{T} = \{1...t...T\}$, where $T$ and $t$ represent the total number of time-steps and a particular time-step in a decision horizon, respectively. The decision horizon is 24 hours with a 30-minute resolution, which makes $T=48$ time-steps. Given that each HEM system has one controllable battery system, the MDP consists of the following:

\begin{itemize}
  \item A set of state variables, $s_{t} \in \mathcal{S}$ to represent electricity demand ($s_t^{\mathrm{d}}$), PV output ($s_t^{\mathrm{pv}}$), electricity tariff ($s_t^{\mathrm{p}}$), grid power ($s_t^{\mathrm{g}}$) and battery state of charge (SOC, $s_t^{\mathrm{b}}$);
  \item A decision variable, $x_{t}^{\mathrm{b}} \in \mathcal{X}$ to describe each control action for the battery system, including charging and discharging rates;
  \item Constraints for all control and state variables, denoted as $\bold{s}^{M}$; and
  \item A random variable, $w_t \in \mathcal{W}$ to capture the perturbation information given by non-controllable inputs, such as demand ($w_t^{\mathrm{d}}$) and PV generation ($w_t^{\mathrm{pv}}$).
\end{itemize}

Thus, the state transition function for describing the evolution of a state from $t$ to $t+1$ is:

\begin{equation} \label{eq1}
	\begin{aligned}
        s_{t+1} = \bold{s}^{M}(s_{t},x_{t}, w_t).
	\end{aligned}
\end{equation} 

Each state contains the information that is necessary and sufficient to make decisions and compute rewards, costs and transitions. An optimal control action is taken to minimize the electricity cost for each observed customer. The problem is solved using DP, which computes the value function that provides the expected future discounted electricity cost for each state. An optimal policy, $\pi$ is extracted from the value function by selecting the state transitions that follow a minimum value function path, such action minimizes the expected sum of future costs over the decision horizon; that is:
\begin{equation} \label{eq2}
	\begin{aligned}
        F = \min_{\pi}\ \mathop{\mathbb{E}} \left(  {\sum_{t=0}^{T} C_{t}(s_t, x_t, w_t)} \right),
	\end{aligned}
\end{equation} 
\noindent where $C_{t}(s_t, x_t, w_t) = s_t^{\mathrm{p}}(s_{t}^{\mathrm{d}}+w_t^{\mathrm{d}}-\eta^{\mathrm{i}}x_t^{\mathrm{i}})$ is the cost of energy incurred at time-step $t$, which accumulates over time. $\eta^{\mathrm{i}}$ is the inverter efficiency ($1/\eta^{\mathrm{i}}$ when inverter power is negative), and $x_{t}^{\mathrm{i}}$ is the inverter power, which is formulated as:
\begin{equation} \label{eq3}
	\begin{aligned}
        x_{t}^{\mathrm{i}} = s_t^{\mathrm{pv}} + w_t^{\mathrm{pv}} - \mu^{\mathrm{b}}x_t^{\mathrm{b}}.
	\end{aligned}
\end{equation} 
\noindent where $\eta^{\mathrm{b}}$ is the battery charging efficiency ($1/\eta^{\mathrm{b}}$ for discharging efficiency). The energy balance constraint \eqref{eq4}, and battery operation constraints \eqref{eq5} to \eqref{eq8} are shown as follows:

\begin{equation} \label{eq4}
	\begin{aligned}
        s_t^{\mathrm{d}} + w_t^{\mathrm{d}} = \eta^{\mathrm{i}}x_{t}^{\mathrm{i}} + s_{t}^{\mathrm{g}}.
	\end{aligned}
\end{equation} 
\begin{equation} \label{eq5}
	s_{t+1}^{\mathrm{b}} = s_t^{\mathrm{b}} + \big(\eta^{\mathrm{b}} x^{\mathrm{b+}}_{t} - \frac{1}{\eta^{\mathrm{b}}}x^{\mathrm{b-}}_{t}\big).
\end{equation}
\begin{equation}\label{eq6}
	0 \leq x^{\mathrm{b+}}_{t} \leq \gamma^{\mathrm{c}}.
\end{equation}
\begin{equation}\label{eq7}
	0 \leq x^{\mathrm{b-}}_{t} \leq {\gamma}^{\mathrm{d}}.
\end{equation}
\begin{equation}\label{eq8}
	{s}_{\mathrm{min}}^{\mathrm{b}} \leq s_t^{\mathrm{b}} \leq s_{\mathrm{max}}^{\mathrm{b}}.
\end{equation}

The SOC at time-step ${t+1}$ of a battery (${s_t^{\mathrm{b}}}$) is a function of the SOC at time-step ${t}$, and the charging (${x^{\mathrm{b+}}_{t}}$) and discharging (${x^{\mathrm{b-}}_{t}}$) rates for this time interval, given by \eqref{eq5}. The charging and discharging rates are constrained by the maximum charging and discharging rates, denoted $\gamma^{\mathrm{c}}$ and $\gamma^{\mathrm{d}}$, respectively, given by \eqref{eq6} and \eqref{eq7}. In addition, the state of charge cannot exceed the minimum and maximum SOC at all times, as described in \eqref{eq8}. For more details regarding the HEM formulation, please visit \cite{keerthisinghe2016fast}.   

Solving this HEM problem using DP for each customer for one year requires {3} hours. This is time-consuming, and therefore impractical within MC analysis.    

\subsection{Policy Function Approximation}
MC analysis requires thousands of runs, so solving the HEM problem exactly is computationally prohibitive. Instead we use a PFA, implemented in Step 5 in Module 2. 
The PFA, illustrated in Fig.~\ref{fig1}, refers to a lookup table that returns a battery schedule for a given set of inputs, including PV output, demand, electricity tariff and SOC at the previous time-step. 
The battery scheduling decisions generated in Step 4 in Module 2 are used to train the ANN. 
Similar to the HEM problem, the choice of the ANN is arbitrary. We showed in \cite{Chanaka2018} that several machine learning techniques could be used with similar performance.
In this work, we use a \textit{recurrent neural network} (RNN) because it has been shown to provide close-to-optimal performance when executing battery schedules trained on similar data \cite{Chanaka2018}. The PFA algorithm is shown in Algorithm \ref{alg0}, and explained in more details below.

\begin{algorithm}[t]
	\small
	\caption{PFA Algorithm}
	\label{alg0}
	\begin{algorithmic}[1]
        \STATE Obtain battery scheduling policies by solving the HEM problem for $n \in \mathcal{N}$ over one year.
        \STATE Train an ANN with the population $\mathcal{N}$ by iterating through Steps 3 to 7.
        \FOR{$n \in \mathcal{N}$}
		    \FOR{$t \in \mathcal{T}$}
                \STATE Train an ANN using $s_{t-1}^{\mathrm{b}}$, $s_{t}^{\mathrm{d}}$, $s_{t}^{\mathrm{pv}}$ and $s_{t}^{\mathrm{p}}$ as inputs, and $s_{t}^{\mathrm{b}}$ as target.
		    \ENDFOR
		\ENDFOR
        \STATE Use the trained ANN to compute optimal battery schedules for sythesized customers by iterating through Steps 9 to 15.
        \FOR{$m \in \mathcal{M}$}
		    \FOR{$t \in \mathcal{T}$}
                \STATE Compute $s_t^{\mathrm{b}}$ from the ANN.
                \STATE Modify $x_t^{\mathrm{b}}$ and $s_t^{\mathrm{b}}$ so that they are within the constraints \eqref{eq6} to \eqref{eq8}.
                \STATE $s_t^{\mathrm{b}} = s_{t+1}^{\mathrm{b}}$. 
		    \ENDFOR
		\ENDFOR
	\end{algorithmic}
\end{algorithm}

\tikzstyle{blocks} = [rectangle, draw, very thick, text width=5.6em, text centered, rounded corners, minimum height=6em]
\tikzstyle{lines} = [draw, -latex']
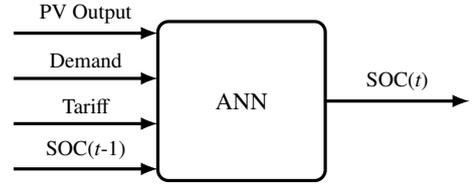
\begin{figure}[t]
	\begin{center}
		\begin{tikzpicture}
		\node [blocks] (input) {\small{ANN}};
		\draw[-latex, black, very thick] (-3,0.9)--(-1.1,0.9) node[pos=0.5,above] {\footnotesize{PV Output}};
		\draw[-latex, black, very thick] (-3,0.3)--(-1.1,0.3) node[pos=0.5,above] {\footnotesize{Demand}};
		\draw[-latex, black, very thick] (-3,-0.3)--(-1.1,-0.3) node[pos=0.5,above] {\footnotesize{Tariff}};
		\draw[-latex, black, very thick] (-3,-0.9)--(-1.1,-0.9) node[pos=0.5,above] {\footnotesize{SOC(\footnotesize\textit{t}-1)}};
		\draw[-latex, black, very thick] (1.1,0)--(3,0) node[pos=0.5,above] {\footnotesize{SOC(\footnotesize\textit{t})}};
		\end{tikzpicture}
		\caption{The PFA model.}
		\label{fig1}
	\end{center}
\end{figure}

The training process requires a training data set, which includes historical demand ($s_{t}^{\mathrm{d}}$), PV generation ($s_{t}^{\mathrm{pv}}$), electricity tariff ($s_{t}^{\mathrm{p}}$), and the calculated SOC with a delay of 1 time-step ($s_{t-1}^{\mathrm{b}}$), while the SOC at the current time-step ($s_{t}^{\mathrm{b}}$) is the target. The ANN learns to use the present and the most recent information to predict the output (target) for each time-step. We use the trained ANN as the PFA to emulate the outputs from the battery scheduling optimization. 

Specifically, we feed the net load traces synthesized in Module 2 into the PFA to compute $s_{t}^{\mathrm{b}}$. To prevent the outputs from violating the battery operation constraints, a control strategy is implemented. This strategy compares the outputs of the PFA at $t$ and $t-1$, and adjust $s_{t}^{\mathrm{b}}$ so that both $s_{t}^{\mathrm{b}}$ and $x_{t}^{\mathrm{b}}$ satisfy the boundaries set by \eqref{eq6} to \eqref{eq8}, before feeding back to the PFA. By doing so, the constraints on both battery operation and capacity are included within the PFA algorithm.

\begin{figure}[t]
	\centering
	\includegraphics[width=\linewidth,keepaspectratio]{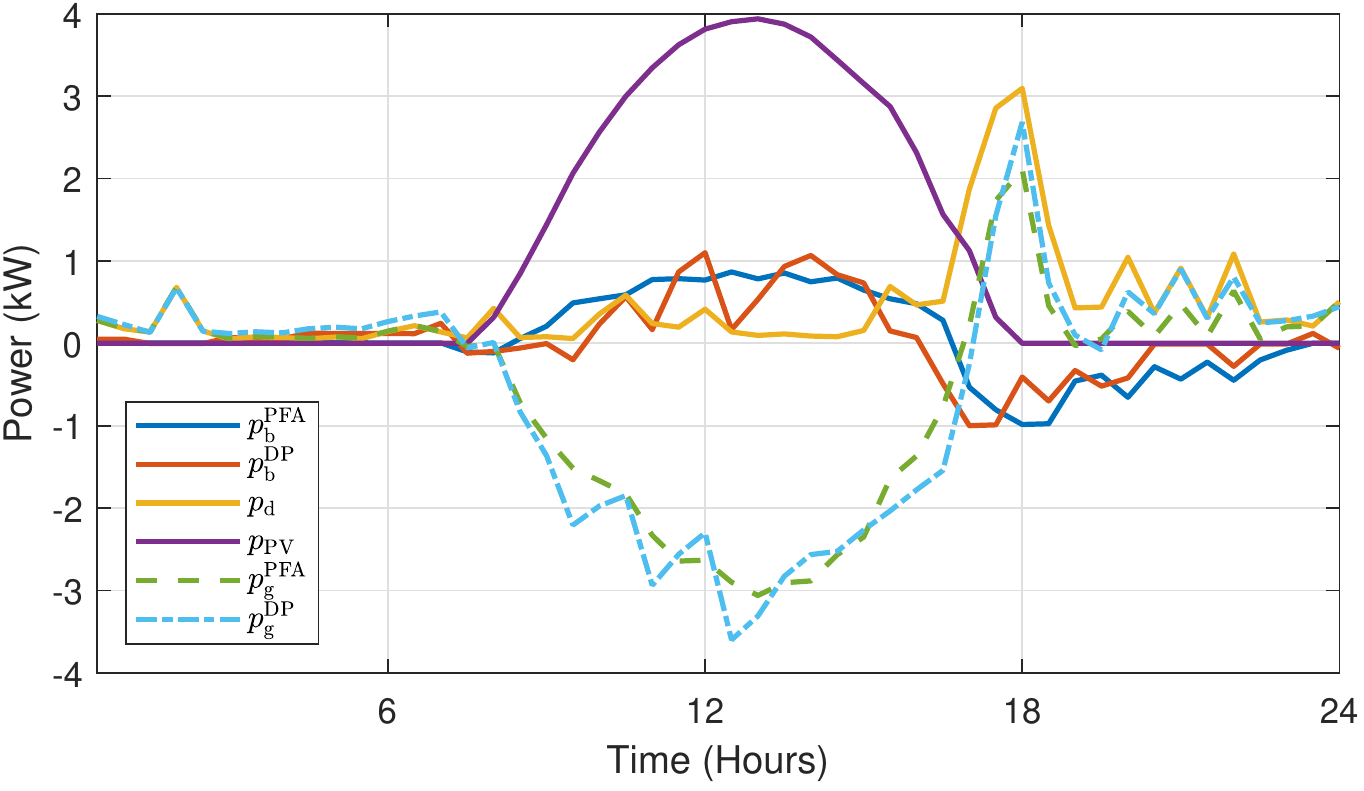}	
    \caption{Example scheduling estimates from PFA (${p^{\mathrm{PFA}}_{\mathrm{b}}}$), the calculated schedules from DP (${p^{\mathrm{DP}}_{\mathrm{b}}}$), and the corresponding PV generation (${p_{\mathrm{PV}}}$), demand (${p_{\mathrm{d}}}$), grid powers, ${p^{\mathrm{PFA}}_{\mathrm{g}}}$ and ${p^{\mathrm{DP}}_{\mathrm{g}}}$, from PFA and DP, respectively.}
    \label{fig0}
\end{figure}

A set of generated demand and PV profiles were used to verify the accuracy of the PFA before feeding the net load traces to the ANN (Step 6 in Module 3). Fig.~\ref{fig0} illustrates the difference between the calculated and estimated battery schedules from DP and the PFA, respectively, for one particular day. Specifically, the energy costs for the same customer are {\$}{2.24} and {\$}{2.29}, using the calculated and estimated battery schedules, respectively. These results are acceptably close, hence, we consider the loss in performance of the PFA to be fit for purpose. This result corroborates those in \cite{Chanaka2018} that the PFA provides close-to-optimal battery schedules.

\section{Probabilistic Impact Assessment Framework} \label{section4}
To probabilistically assess the impact of residential batteries on distribution networks, Module 3 incorporates the HEM problem within the MC analysis, summarized in Algorithm \ref{alg1}. The MC simulation is run 100 times for 11 PV and 3 battery penetration levels, resulting in 3300 yearly power flow simulations with a half hourly resolution. 
The results provide insights regarding the probabilities for a technical issue to occur based on different PV and battery penetration levels, which define the percentage of customers that have a PV system alone, or a PV-battery system. 

\subsection{Sampling Process}
The Ausgrid data set \cite{SHED} provides smart meter and PV generation data with a 30-minute resolution for 150 customers for a period of three years. 
We increased the size of individual PV systems to better represent the average residential PV size, which was around \SI{5.5}{kW} in Australia in 2017.
The battery size is decided based on the size of each PV system. In Australia, \SI{2}{kWh} of battery is typically used per \SI{1}{kW} of PV installed. The batteries used are from LG and Tesla, which provide three battery sizes to match the PV size ranges. The detailed allocation is summarized in Table \ref{T1}.
\begin{table}[t]
    \footnotesize
	\centering
	\caption{Battery storage specifications}
	\begin{tabular}{lllllll}
		\\[-1.8ex]\hline 
		\hline  
		Attached PV size (kW)    & $\leq4$     &5-6         &7-10     \\
		Battery Capacity (kWh) &6.5        &9.8        &14.0          \\
		Battery Power (kW) 	 &4.2        &5.0        &5.0  \\ 		
		Manufacturer         & LG & LG & Tesla 		\\ 
		\hline 
		\hline \\[-1.8ex] 		  
	\end{tabular}
	\label{T1}
\end{table}


For each customer obtained from the Ausgrid smart meter data set, we calculated the battery scheduling policies for the whole year using DP, which we then used to train the ANN in the PFA. The PFA was then used to provide battery scheduling policies for the 3000 synthetic demand and PV traces generated in Module 1. 
To capture the uncertainties in the power flow study, we probabilistically sample from this pool of synthetic net load traces for random allocation of loads, PV and battery systems.
In more detail, each load assignment accounts for eleven levels of PV penetration, denoted ${P_\mathrm{PV}}$, ranging from \SI{0}{\%} to \SI{100}{\%}. Specific to each ${P_\mathrm{PV}}$, a set of load traces with a PV system ${\mathcal{D}_\mathrm{PV}}$ is randomly sampled from the set of load traces ${\mathcal{D}_\mathrm{load}}$. Following this, three battery penetration levels  ${P_\mathrm{b}}$ (\SI{0}{\%}, \SI{50}{\%}, \SI{100}{\%}) are implemented for each ${P_\mathrm{PV}}$. A set of traces with a battery system ${\mathcal{D}_\mathrm{b}}$ is randomly drawn from ${\mathcal{D}_\mathrm{PV}}$. This process covers both Steps 7 and 8 in Module 3. 

\begin{algorithm}[t]
	\small
	\caption{Probabilistic Impact Assessment}
	\label{alg1}
	\begin{algorithmic}[1]
		\FOR{i = 1:100}
		\STATE Randomly sample net load traces from ${\mathcal{D}_\mathrm{load}}$.
		\FOR{$P_{\mathrm{PV}}$ = $0{\%}$:$10{\%}$:$100{\%}$}
		\STATE Randomly sample from ${\mathcal{D}_\mathrm{load}}$ a set of PV assignments ${\mathcal{D}_\mathrm{PV}}$.
		\FOR{$P_{\mathrm{b}}$ = $0{\%}$:$50{\%}$:$100{\%}$}
		\STATE Randomly sample from ${\mathcal{D}_\mathrm{PV}}$ a set of battery assignments ${\mathcal{D}_\mathrm{b}}$.
		\STATE Run power flow analysis.
		\ENDFOR
		\ENDFOR
		\ENDFOR
	\end{algorithmic}
\end{algorithm}

\subsection{Power Flow Analysis}
The sampled data are used to run yearly power flow simulations for all MC realization paths (Step 9 in Module 3)\footnote{We use OpenDSS \cite{opendss} for power flow analysis.}.
Yearly voltage profiles for each customer and feeder head loading are used to determine the probabilities of a technical problem, namely over-voltage and/or congestion problem, according to the specific metrics defined below.

\subsubsection{Voltage Problem}
The maximum and minimum phase voltage thresholds at each busbar are \SI{241.5}{\V} (1.05 pu) and \SI{218.5}{\V} (0.95 pu) phase-to-neutral, respectively. This provides room for voltage rise and drop when peak load or high PV penetration occurs. The daily voltage profile is calculated for each customer and checked for compliance with the modified standard BS EN 50160 \cite{BSI}, which states that customers' voltages must be between 0.95 and 1.05 p.u during 95\% of the time and never below 0.9 or above 1.1 p.u.

\subsubsection{Thermal Loading Problem}
The thermal loading level is defined by the ratio of the half-hourly maximum current to the transformer capacity. Specifically, if the ratio is greater than 1, the network has a thermal problem. 

\subsubsection{Phase unbalance}
This study also investigates the effects of PV-battery systems on the voltage unbalance factor, which is a measure of the phase unbalance \cite{pillay2001definitions}. Specifically, it is the ratio between negative and positive sequence voltages. 

\section{Results and Evaluation} \label{section5}
The probabilistic impact assessment framework is applied to two typical Australian LV networks and one UK LV network. The results are analyzed by running yearly power flow analysis. The magnitude of a technical problem (over-voltage, transformer loading level and phase unbalance) is recorded at different levels of ${P_\mathrm{PV}}$ and ${P_\mathrm{b}}$.  

\subsection{Computational Performance}
Computational performance is the linchpin of the proposed framework. Solving the scheduling problem for each net load trace for one year using DP requires {3} hours. Given the 3300 yearly power flow simulations (100 MC runs for 11 PV penetration levels each with three battery penetration levels), this is clearly impractical. By using the PFA algorithm to emulate the battery scheduling policies, we were able to reduce the computational time to five minutes for each customer, which is more than a \SI{95}{\%} reduction. Training the ANN only takes  30 minutes, which is negligible. Thus, the use of PFAs is essential to making the entire MC process computationally feasible.

\begin{table}[t]
    \footnotesize
	\centering
	\caption{LV test feeders}
	\begin{tabular}{c@{\hspace{0.16cm}}c@{\hspace{0.16cm}}c@{\hspace{0.16cm}}c@{\hspace{0.16cm}}c@{\hspace{0.16cm}}c@{\hspace{0.16cm}}c@{\hspace{0.16cm}}}
		\hline 
		\hline 
		Feeder Name      & Length (m)   &No. of customers &Feeder head ampacity (A) \\
		AUS 1  		&10235       &302        &1155       	\\
		AUS 2 	    &5656       &223        &1200			\\ 	
        UK  		&5656       &223        &400			\\
        \hline 
		\hline 	  
	\end{tabular}

	\label{T2}
\end{table}

\begin{figure*}[t]%
    \centering
    \includegraphics[width=4cm,keepaspectratio]{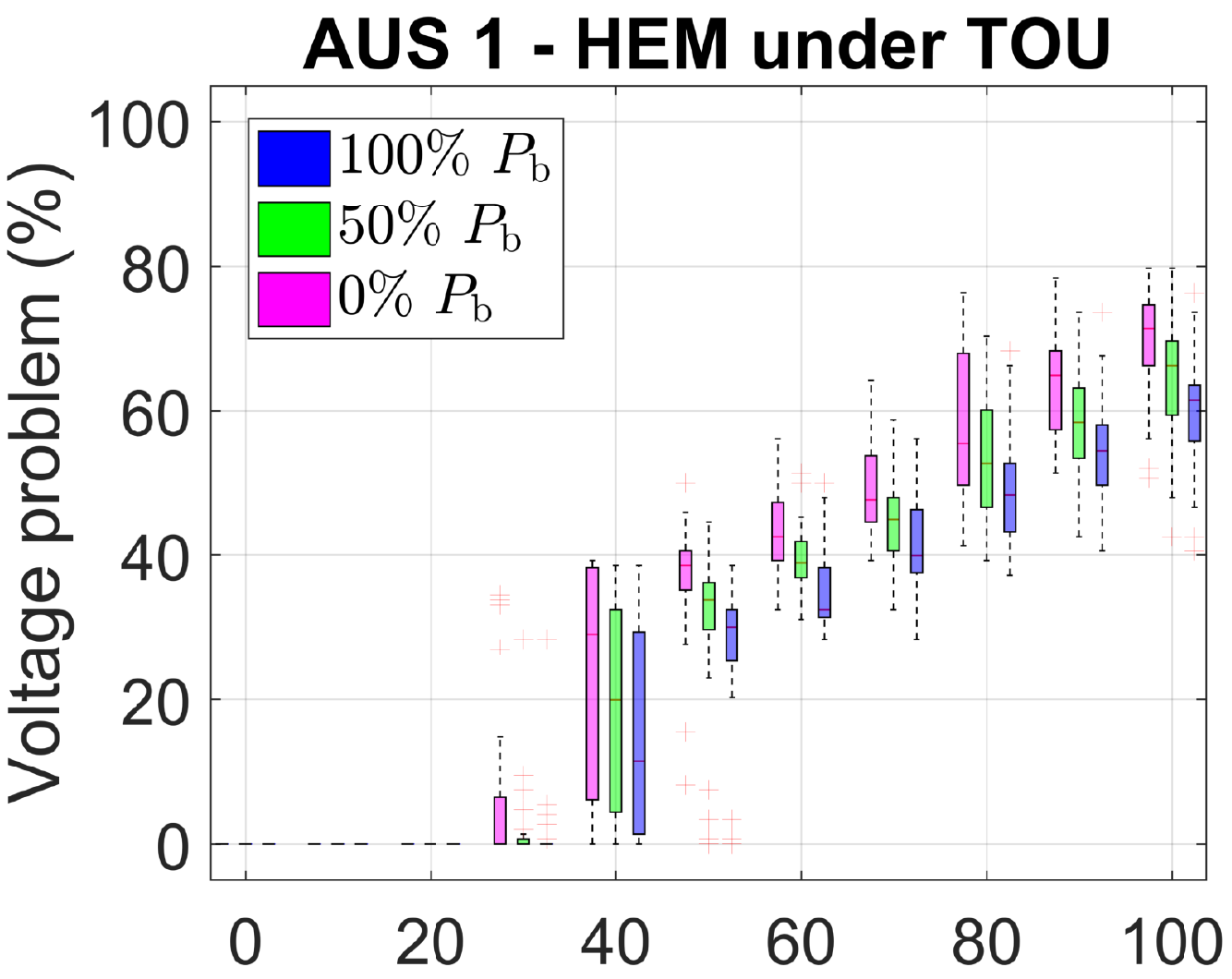}%
    \qquad
    \includegraphics[width=4cm,keepaspectratio]{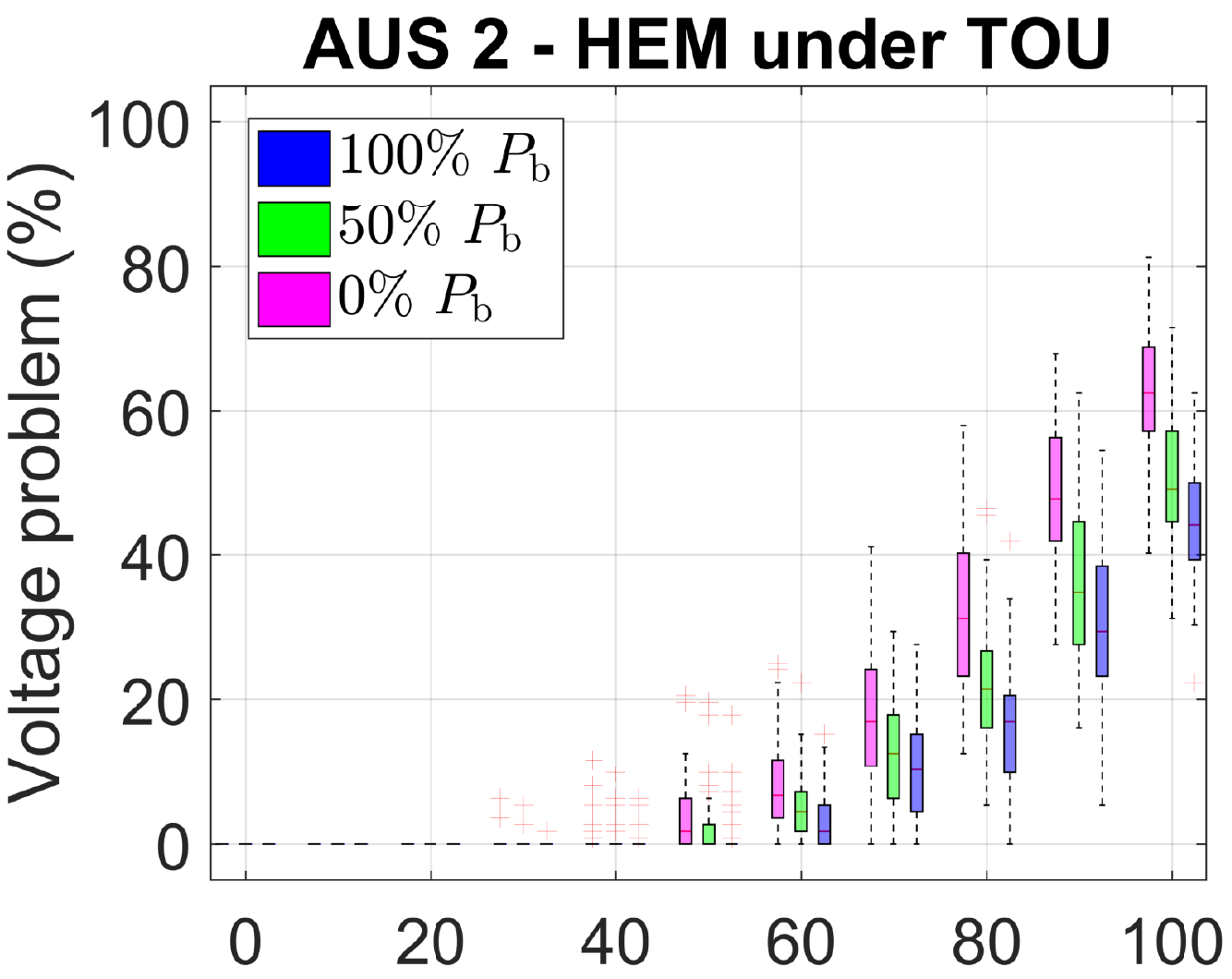}%
    \qquad
    \includegraphics[width=4cm,keepaspectratio]{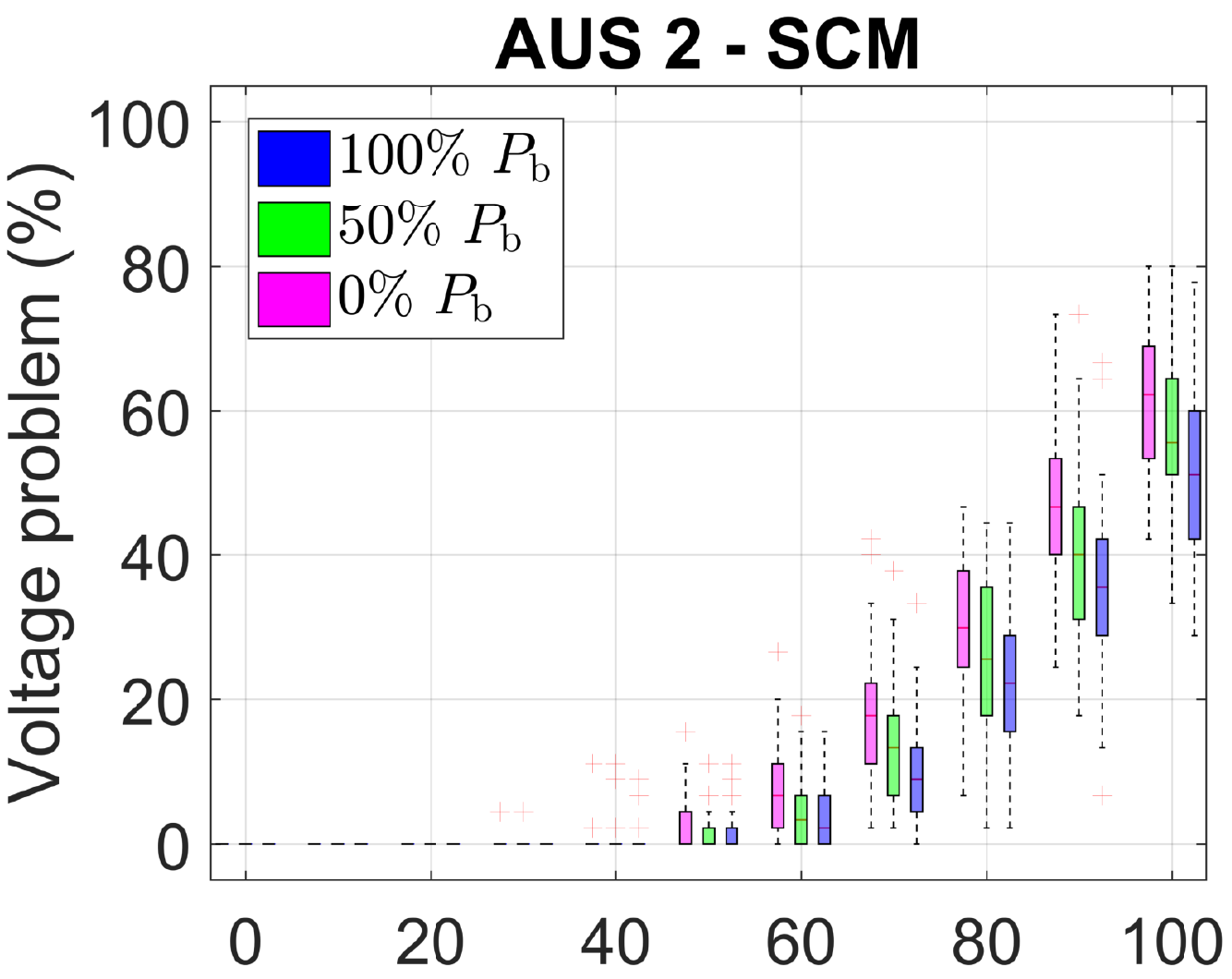}%
    \qquad
    \includegraphics[width=4cm,keepaspectratio]{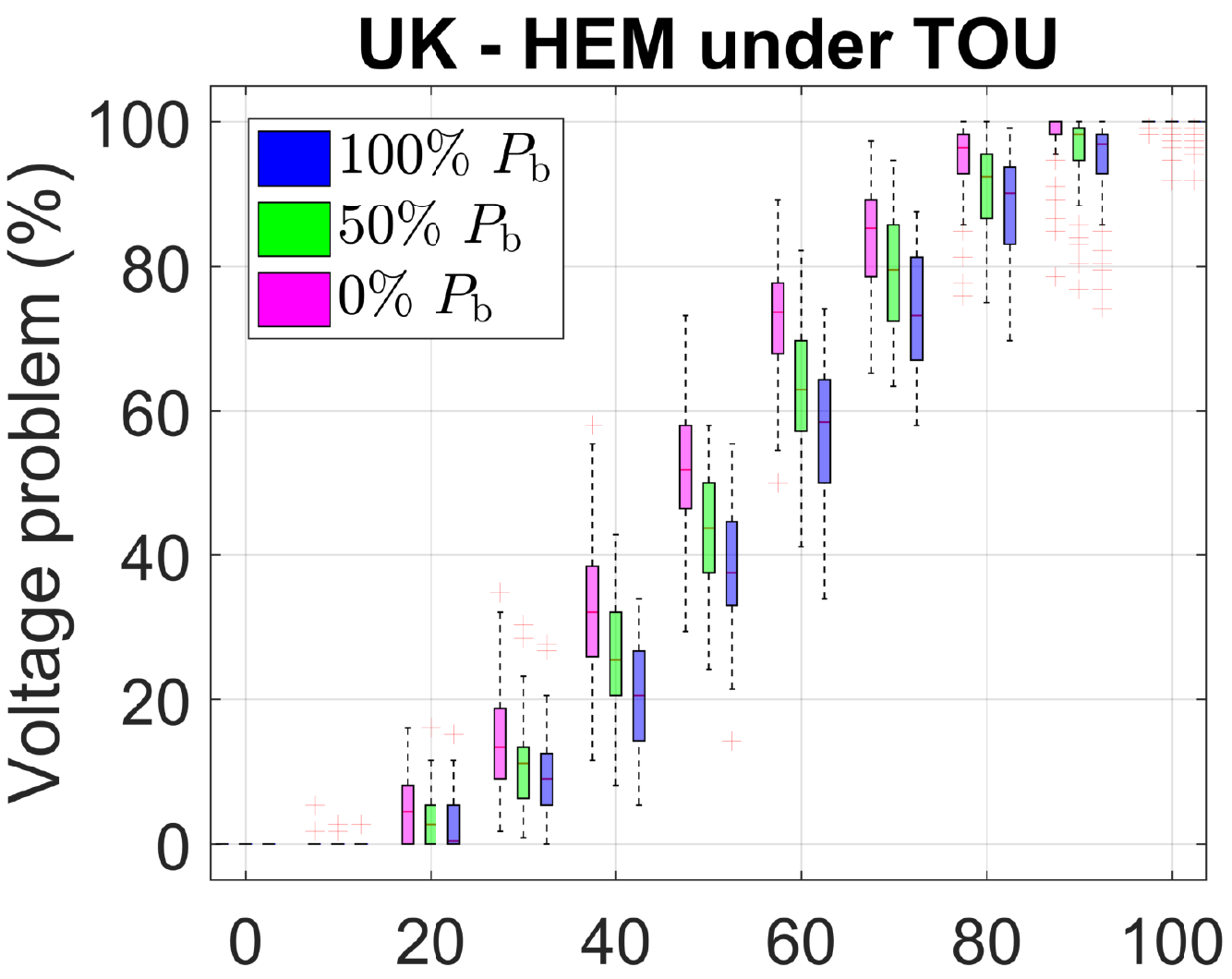}%
    \qquad
    \includegraphics[width=4cm,keepaspectratio]{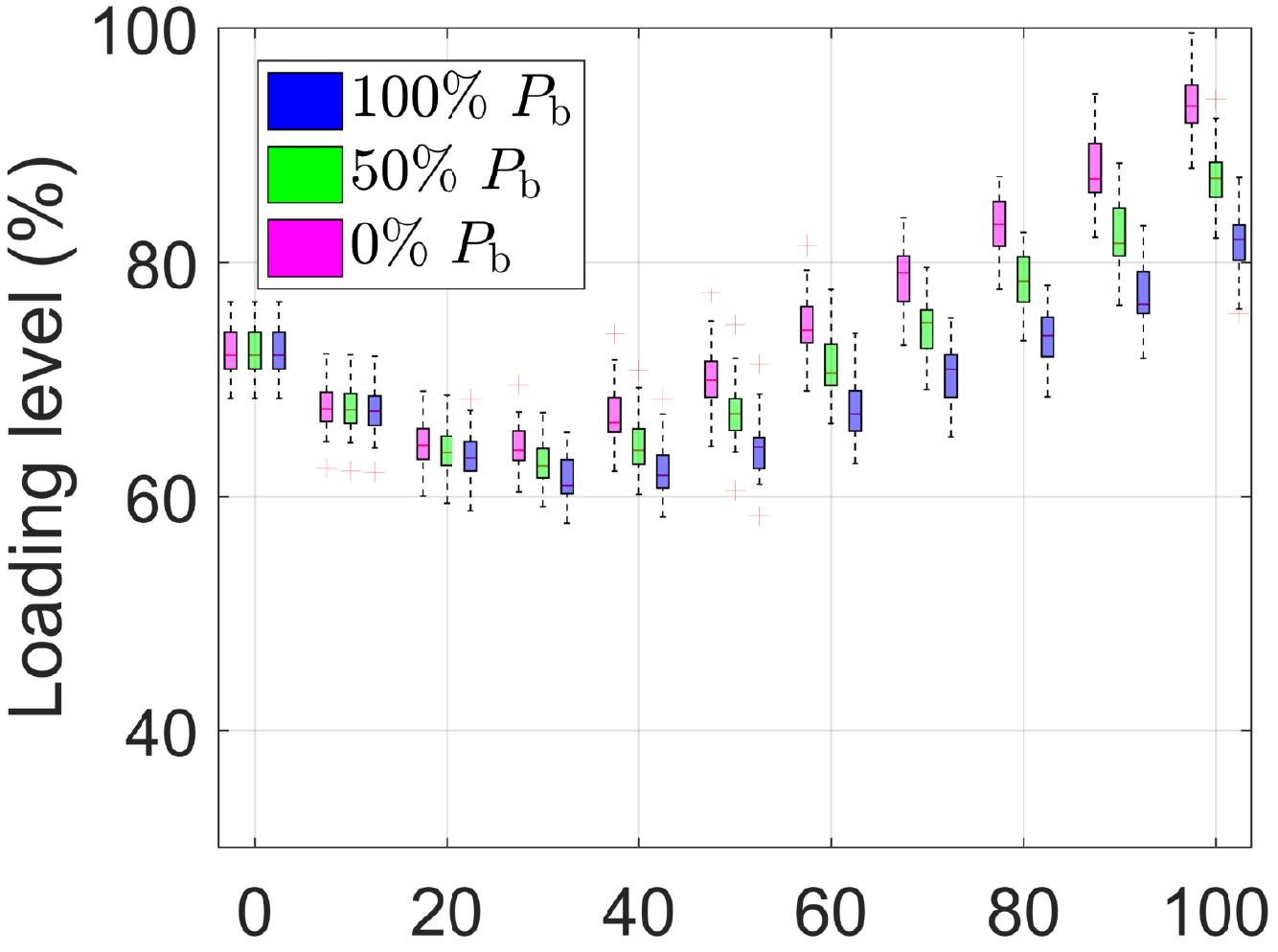}%
    \qquad
    \includegraphics[width=4cm,keepaspectratio]{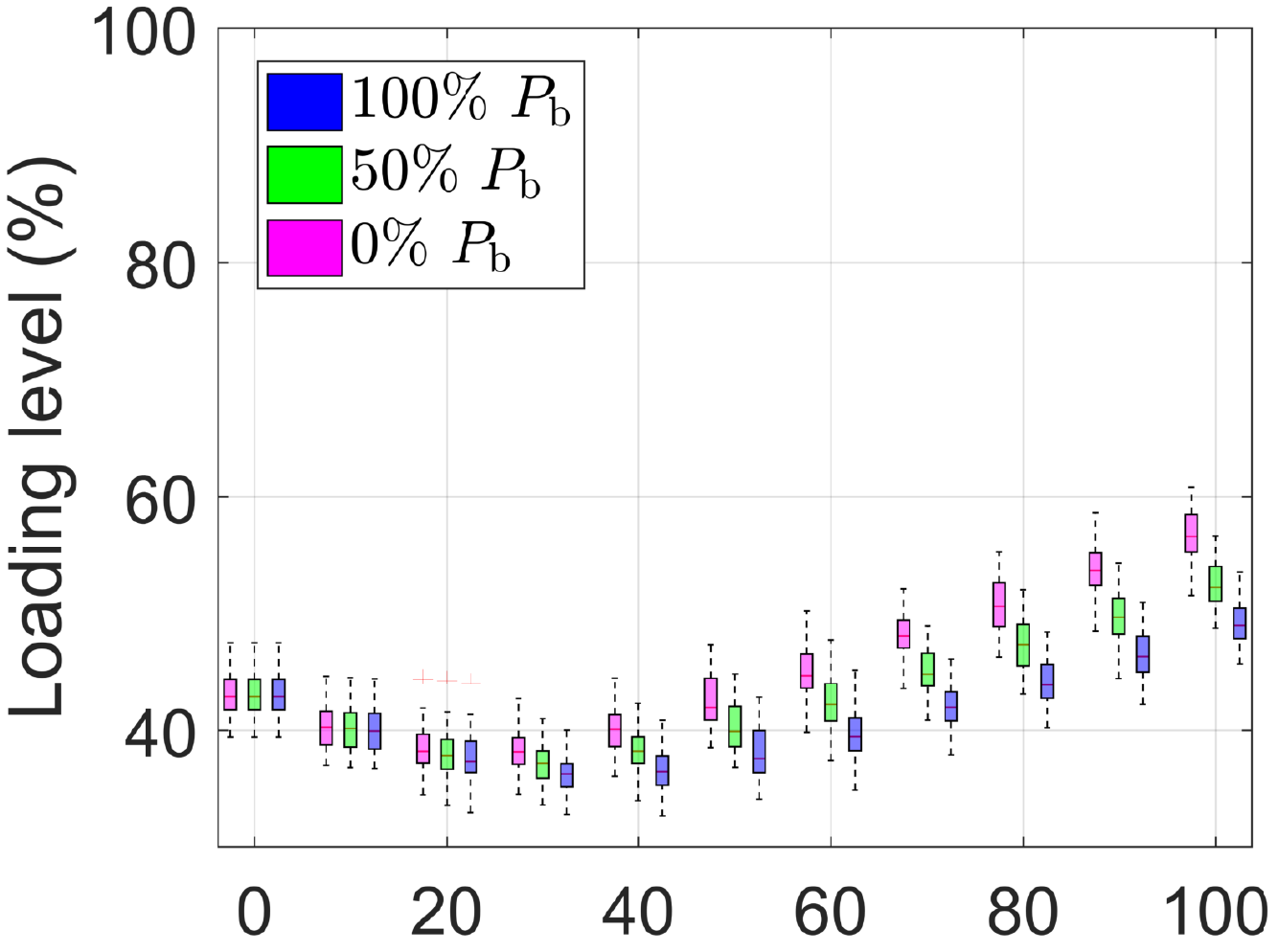}%
    \qquad
    \includegraphics[width=4cm,keepaspectratio]{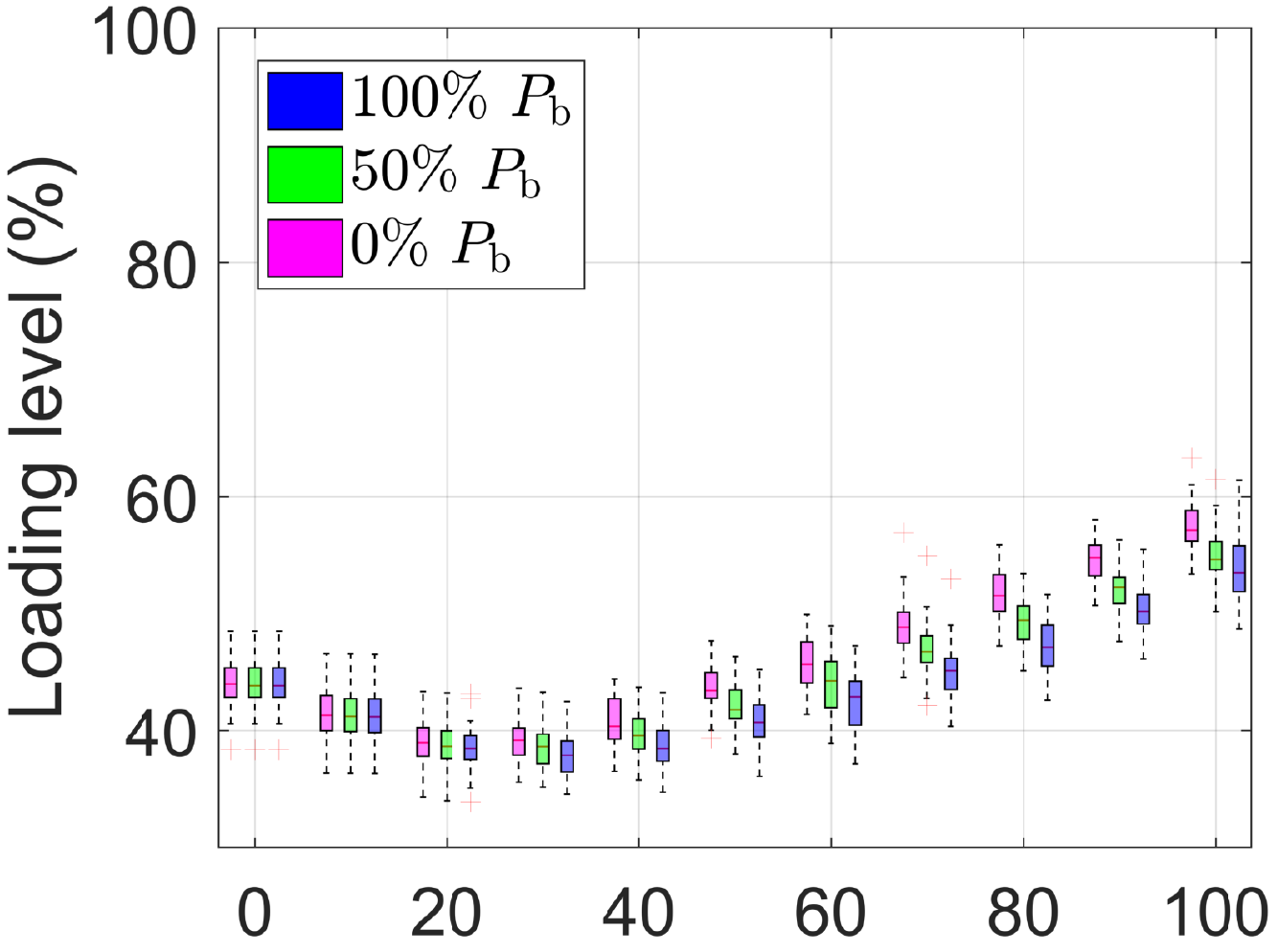}%
    \qquad
    \includegraphics[width=4cm,keepaspectratio]{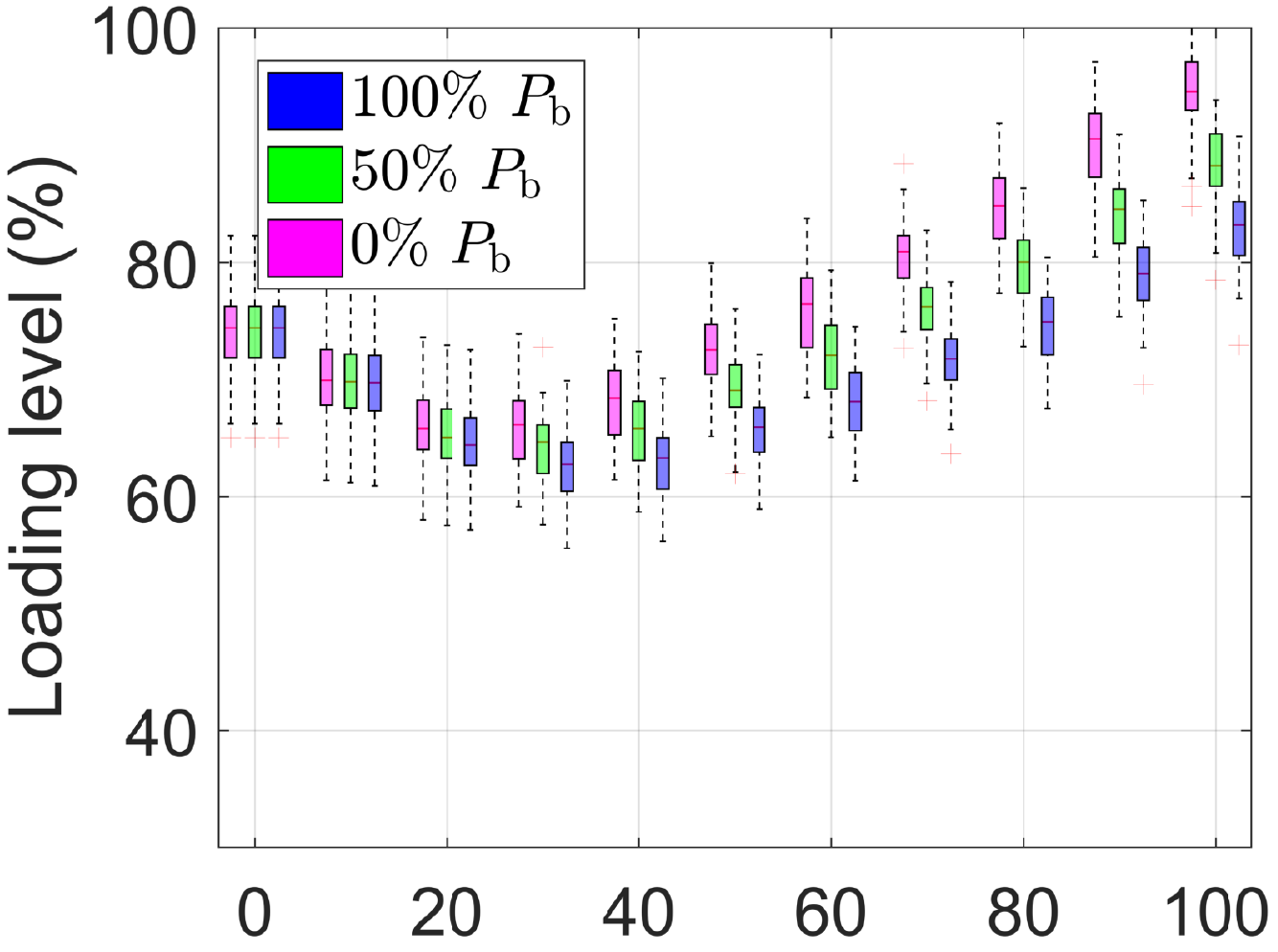}%
    \qquad
    \includegraphics[width=4cm,keepaspectratio]{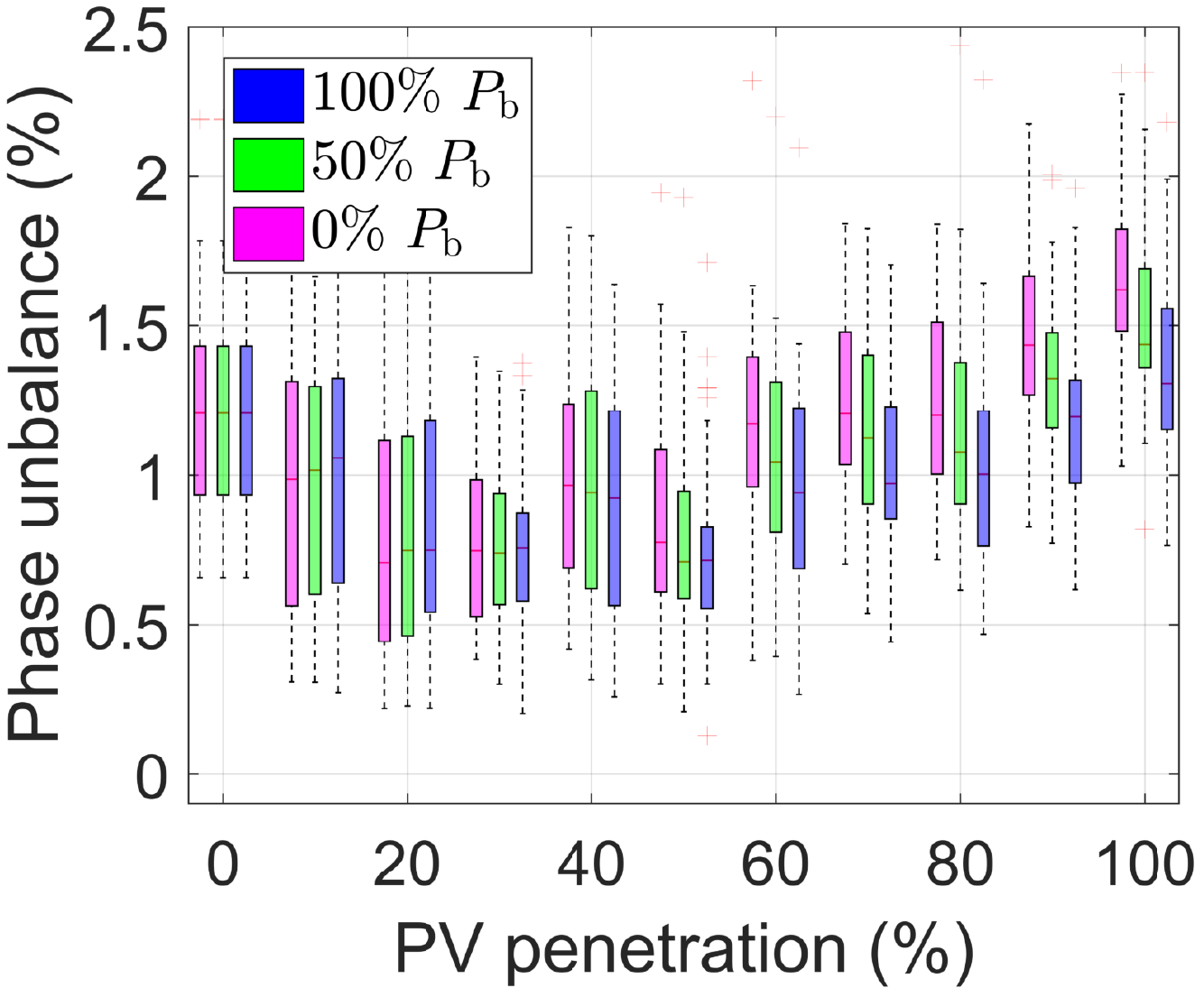}%
    \qquad
    \includegraphics[width=4cm,keepaspectratio]{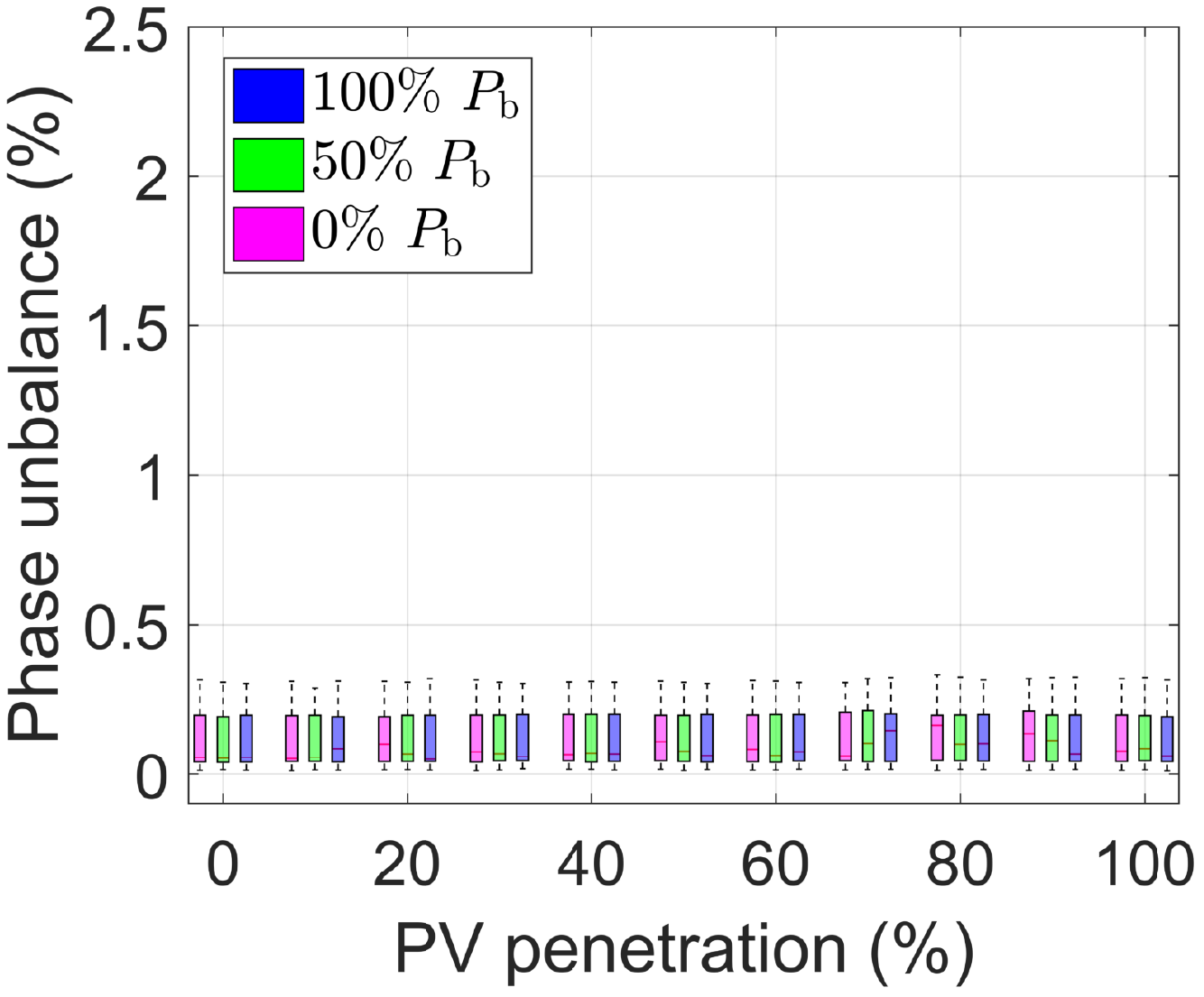}%
    \qquad
    \includegraphics[width=4cm,keepaspectratio]{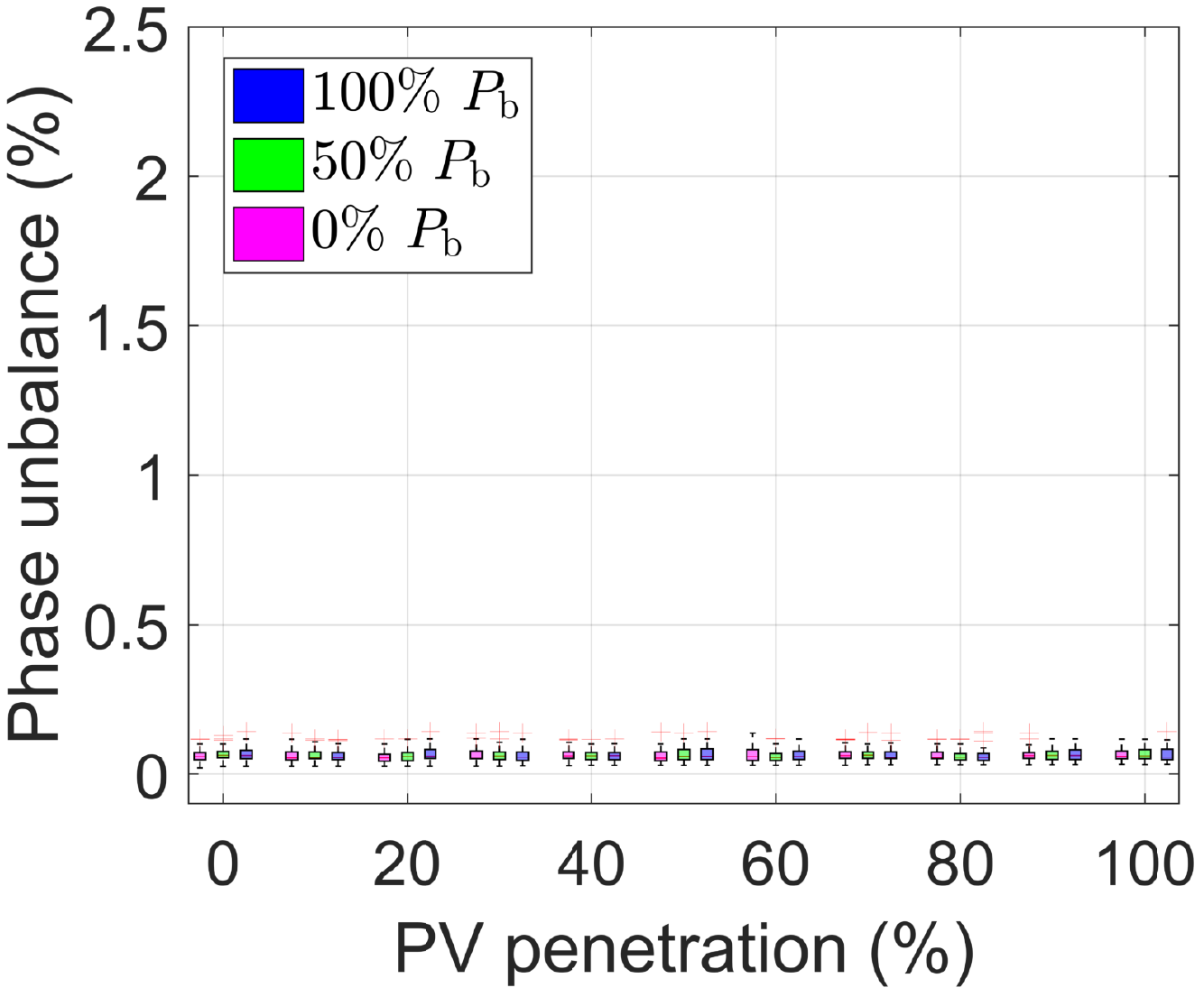}%
    \qquad
    \includegraphics[width=4cm,keepaspectratio]{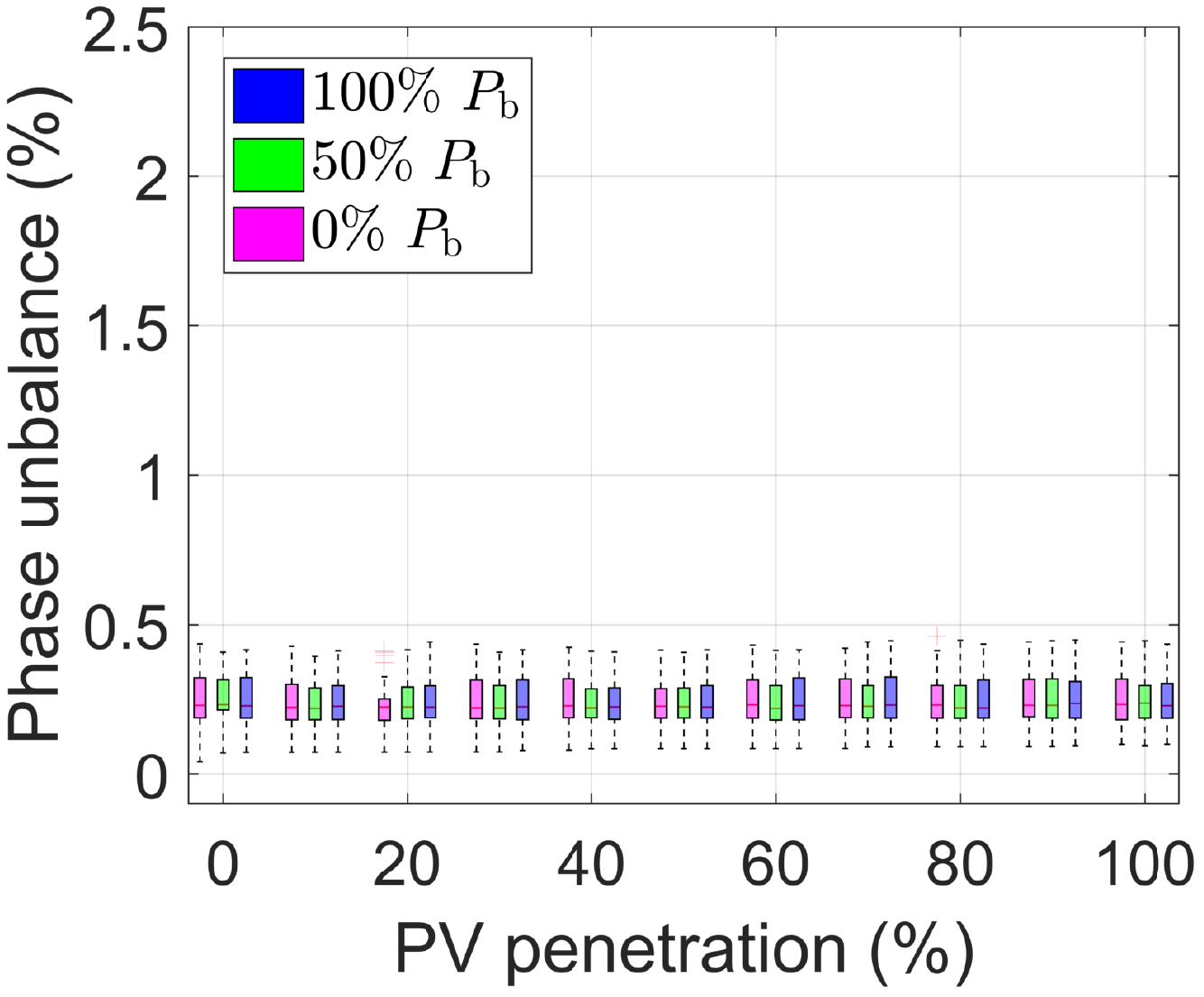}%
	\caption{Percentage of customers with voltage problems, transformer loading level and phase unbalance. Pink, green and blue bars represent \SI{0}{\%}, \SI{50}{\%} and \SI{100}{\%} battery penetration levels, respectively. Each bar from top to bottom shows the maximum, {75} percentile, median, {25} percentile and minimum value.}
	\label{fig3}
\end{figure*}

\subsection{Test Networks}
In order to evaluate the method, two four-wire three-phase unbalanced LV test networks with different lengths are adopted from Electricity North West Limited (ENWL), a British network operator \cite{enwlreport}. Typically, Australian LV networks are designed to have higher capacity than the UK ones, mainly due to much larger air-conditioning loads. To match this design, the UK test networks are transformed into Australian-type LV networks by tripling the transformer and line capacity\footnote{For transformers, we reduced the impedance, while for transmission lines we only reduced the resistance. The reactance mainly depends on the distance between the conductors, so we left it unchanged.}. These test feeders are denoted as AUS 1 and AUS 2, respectively. Each feeder is supplied by a \SI{2250}{kVA} \SI{11}{kV}/\SI{0.4}{kV} 3-phase transformer. In addition, one of the selected UK feeders is supplied as the third test case, denoted UK, with a lower feeder head ampacity for comparison. The details of the test networks are summarized in Table~\ref{T2}. We sample from the pool of net load traces synthesized in Module 1 of our method for allocation to load points. 

\subsection{HEM formulation}
The HEM objective is to minimize energy expenditure under a \textit{time-of-use} (ToU) tariff with the peak demand period from 2pm--8pm. The PV buy-back rate (the ``feed-in tariff'') is much lower than the electricity tariff so there is no incentive for the HEM to export power to the grid. 
As a benchmark, we use \textit{self-consumption maximization} (SCM) heuristic scheduling strategy, whereby the energy from the solar PV is first used to meet the demand, and then any excess PV generation is used to charge the battery, or exported to the grid if the battery is full. For brevity, the SCM is applied only to AUS 2 which has a larger potential for greater technical problems, and the results are compared with the HEM under ToU tariff. 

\subsection{Voltage Problems}
The frequency of voltage problems with respect to increasing $P_\mathrm{PV}$ and $P_\mathrm{b}$ on the LV feeders is shown in Fig.~\ref{fig3}, row one. The percentage of customers with a voltage problem follows an increasing trend across all test feeders with respect to rising $P_\mathrm{PV}$, especially from \SI{30}{\%} to \SI{100}{\%}, while the UK feeder presents more voltage problems due to higher line impedances. 

\begin{figure}[t]
	\centering
	\includegraphics[width=\linewidth,keepaspectratio]{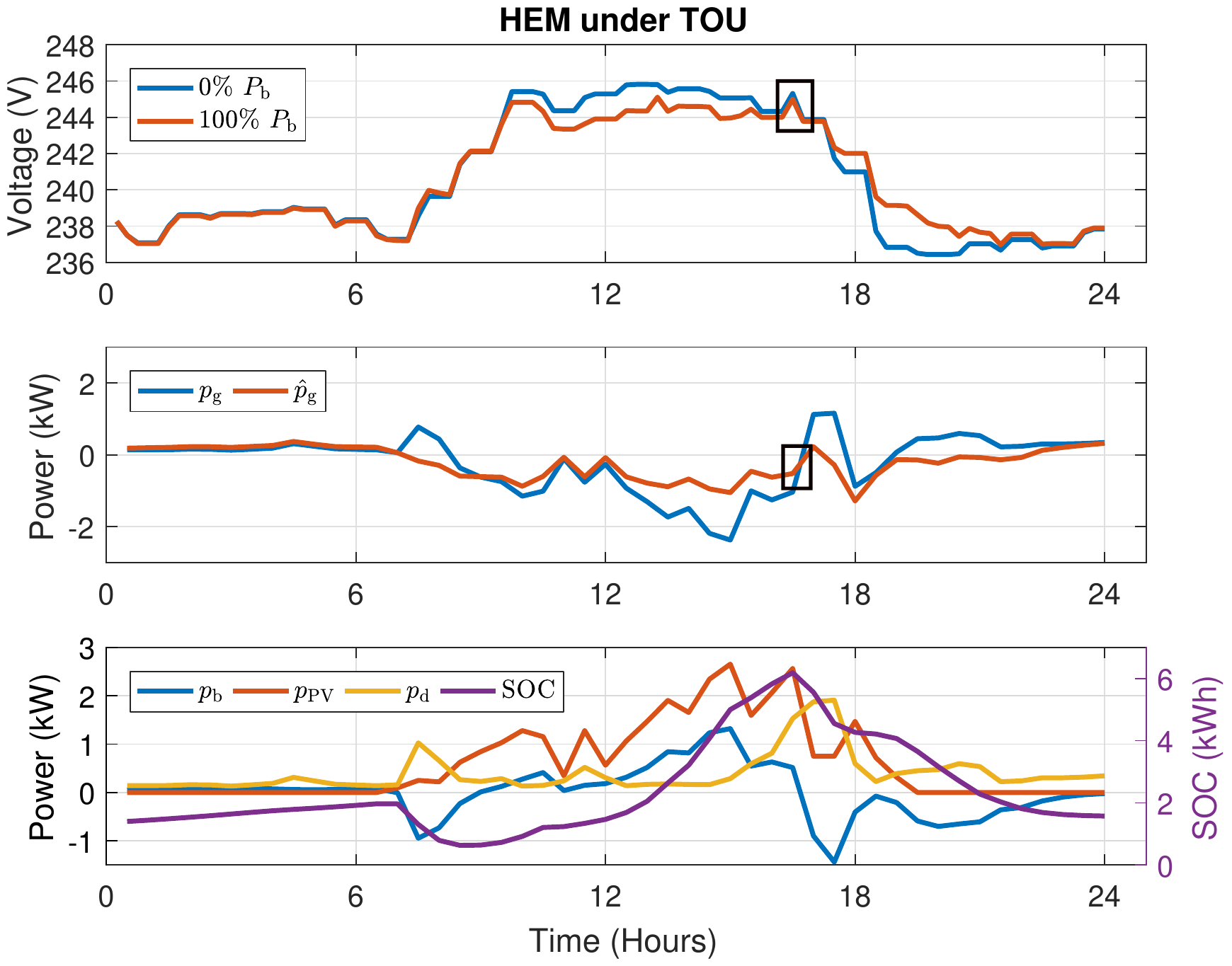}
	\caption{Voltage profiles (top), grid power (${p}_\mathrm{g}$ and $\hat{p}_\mathrm{g}$ denote grid power with and without battery, respectively) (middle), battery scheduling ($p_\mathrm{b}$), PV ($p_\mathrm{PV}$), demand ($p_\mathrm{d}$), and the SOC (bottom) of a customer with \SI{3}{kW} PV and \SI{6.5}{kWh} battery on AUS 2 on a particular summer day, with \si{100}{\%} $P_\mathrm{PV}$.}
	\label{fig4}	
\end{figure}

Voltage problems can be reduced by {10}-\SI{20}{\%} across all test feeders using HEM under ToU (Fig.~\ref{fig3}). This scheduling strategy encourages batteries to charge when electricity price is low, and discharge when the price is high (during peak hours). However, the time-span for high PV outputs can extend and even overlap with peak demand, especially in summer. This is illustrated for some specific case in Fig.~\ref{fig4}, in which the peak demand occurs between 4 and 6pm, causing the battery to discharge during high PV output. This reduces the grid power supply ($p_\mathrm{g}$), when compared to the case without the battery ($\hat{p_\mathrm{g}}$). As a result, $p_\mathrm{g}$ and $\hat{p_\mathrm{g}}$ cross at around 4:30pm, where the voltages become the same (as highlighted in the black boxes). Furthermore, at 4:30pm, rising demand causes the battery to decrease its charging power at high PV output, which keeps the voltage at a high level. In these scenarios, HEM under ToU is less effective at reducing over-voltage problems. It should be also realized that the SOC reaches the peak at 5pm while the battery is still charging. This is because the battery switches to discharging at 5:30pm, which leads to a decreasing SOC from 5 to 5:30pm.   

\begin{figure}[t]
	\centering
	\includegraphics[width=\linewidth,keepaspectratio]{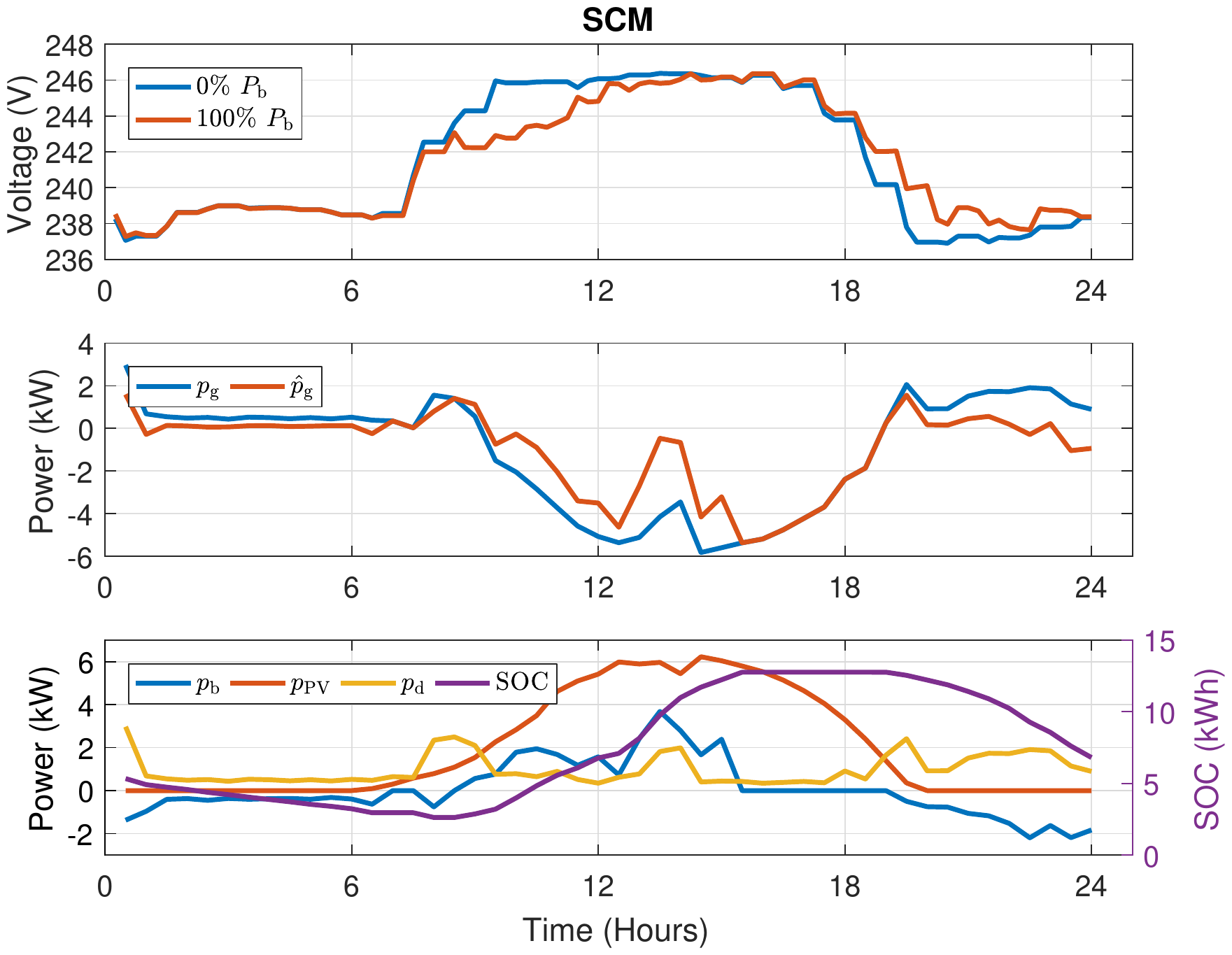}
	\caption{Voltage profiles (top), grid power profiles (middle), battery scheduling, PV, demand and the SOC (bottom) of a customer with \SI{7}{kW} PV and \SI{14}{kWh} battery installed on AUS 2 on a particular summer day, assuming \SI{100}{\%} $P_\mathrm{PV}$.}
	\label{fig5}	
\end{figure}

In addition, longer feeders (AUS 1) experience larger voltage drops, and hence, the rate of increase in frequency of voltage problems with respect to ${P_\mathrm{PV}}$ is lower when compared with the smaller feeders (AUS 2 and UK). This is illustrated in Fig.~\ref{fig3}, where the voltage problems increase at a slower rate towards high $P_\mathrm{PV}$ for AUS 1. More so, there is a greater chance for longer feeders (AUS 1) to concentrate PV installations in particular parts of the network. Consequently, there is a greater variability in the voltage metric at low $P_\mathrm{PV}$, as illustrated at \SI{40}{\%} $P_\mathrm{PV}$ for AUS 1 (Fig.~\ref{fig3}).  

\begin{figure}[t]
	\centering
	\includegraphics[width=\linewidth,keepaspectratio]{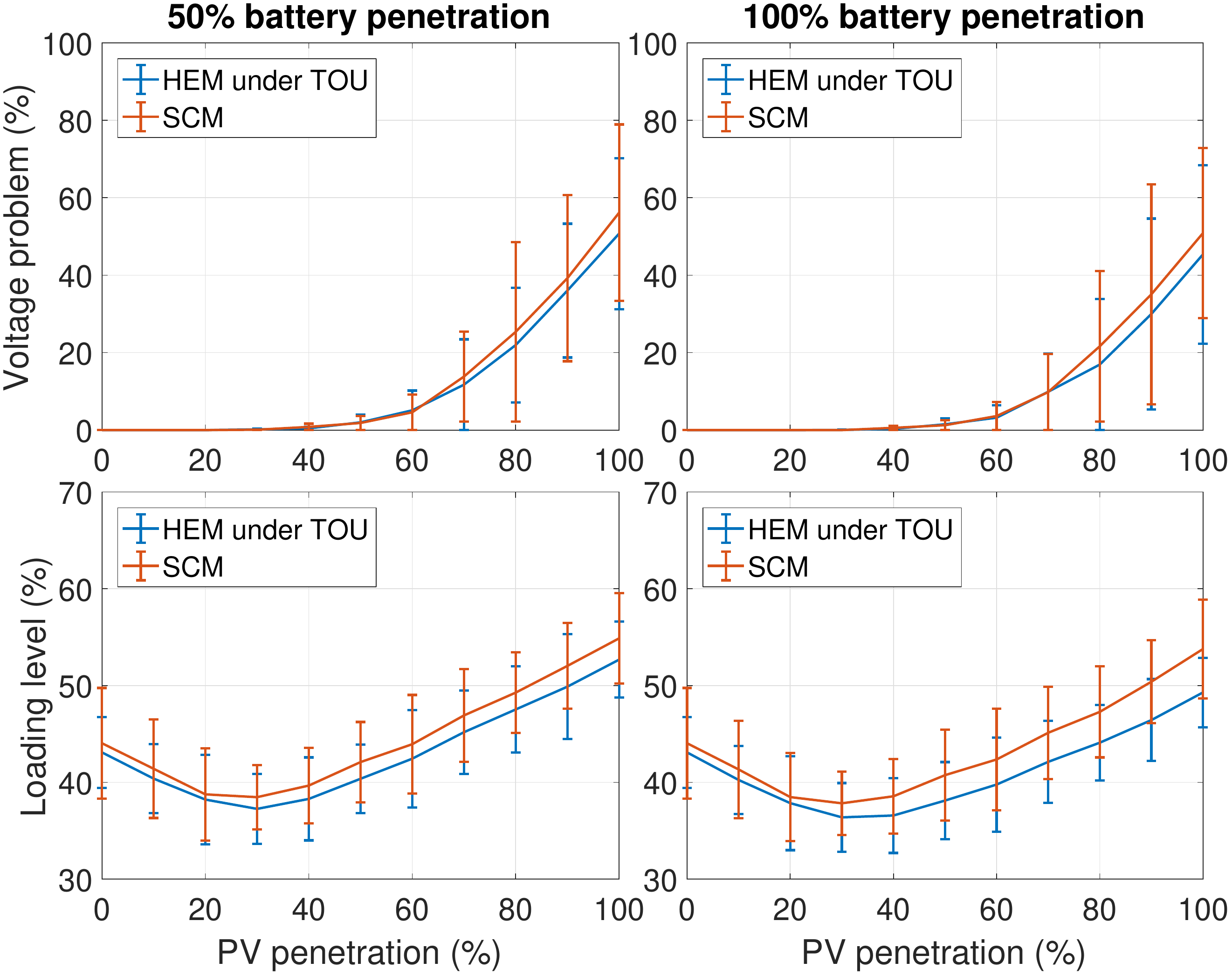}
	\caption{Comparisons between HEM under ToU and SCM on AUS 2.}
	\label{fig6}	
\end{figure}

Compared to the SCM benchmark, HEM under ToU is more effective in mitigating over-voltage problems (Fig.~\ref{fig6}). The SCM forces batteries to charge with excess PV output to reach the battery's full capacity, which usually occurs before the end of the high solar generation time period. Due to this, batteries fail to consistently reduce voltage problems across the entire PV generation period. This can be seen in Fig.~\ref{fig5}, where the battery reaches its full capacity at 3pm (purple curve). As a result, all excess solar generation after 3pm is exported to the grid, as illustrated by the section where $\hat{p_\mathrm{g}}$ overlaps ${p_\mathrm{g}}$, keeping the voltage at a high level between 3 and 6pm. On the other hand, the HEM under ToU considers the price of electricity, which is an indication of the timely demand and PV generation. In this method (Fig.~\ref{fig4}), the charging profile (blue curve in the bottom plot) is more evenly distributed throughout the PV generation period, hence a more consistent reduction of the problems is expected. Although the HEM with ToU helps reduce voltage problems, it is far from a panacea for voltage problems on distribution feeders. 

\subsection{Thermal Problems}
This subsection evaluates the occurrence of thermal problems across all test feeders. The transformer loading drops between \SI{0}{\%} and \SI{40}{\%} $P_\mathrm{PV}$ for all test feeders, as illustrated by Fig.~\ref{fig3}, row two. Within this interval, all solar generation is consumed by demand. However, with greater $P_\mathrm{PV}$ (more than \SI{40}{\%}), excess solar generation is exported to the grid, accumulating at the feeder head and increasing the transformer loading level. All test feeders follow these trends with the turning point at roughly \SI{40}{\%}. Before this point, PV systems alone helps in transformer loading reduction. Additionally, the loading levels are higher for longer feeders (AUS 1), as well as the feeders with lower transformer capacity (UK). 

Batteries reduce the transformer loading levels by charging with solar generation, and then discharge during peak periods. They become more effective as both PV and battery capacities increase after \SI{40}{\%} $P_\mathrm{PV}$.  For example, in AUS 1 (Fig.~\ref{fig3}, row two), a \SI{5}{\%} reduction in thermal loading is achieved at \SI{40}{\%} ${P_\mathrm{PV}}$, this proportion increases to \SI{20}{\%} at \SI{100}{\%} ${P_\mathrm{PV}}$. In contrast to the voltage problem reduction, HEM under ToU is more effective on longer feeders (AUS 1) that have higher battery capacities for charging with excess PV generation. Compared to the benchmark, the HEM under ToU is more effective in thermal loading reduction, as seen in Fig.~\ref{fig6}. This is because the SCM forces the battery to charge to its full capacity before the end of the PV generation period, as explained previously for voltage problem reduction. 

\subsection{Phase Unbalance}
This subsection presents the impacts on the voltage unbalance factor, with the results shown in Fig.~\ref{fig3}, row three. Increasing $P_\mathrm{PV}$ can amplify phase unbalance. Using AUS 1 as an example, when $P_\mathrm{PV}$ on the feeder is low, typically between \SI{0}{\%} and \SI{40}{\%}, solar generation alone helps reduce the phase unbalance. When $P_\mathrm{PV}$ is greater than \SI{40}{\%}, unused solar generation is exported to the grid, and hence, increasing the phase unbalance. In this case, the unbalance is improved by charging the battery with excess solar generation. Specifically, the voltage unbalance factor for AUS 1 is reduced from {1.6}{\%} to {1.2}{\%} at \SI{100}{\%} ${P_\mathrm{PV}}$. PV and battery systems are shown to mitigate cases of high unbalance, as on AUS 1; while the impacts are less pronounced for the other test feeders (AUS 2 and UK) as they are rather balanced to begin with. Overall, the improvement on phase unbalance for either the HEM under ToU or the SCM is limited.

\section{Conclusions} \label{section6}
This paper proposed a novel methodology that can (i) explicitly incorporate battery scheduling in a MC analysis, and (ii) synthesize statistically representative demand and PV profiles using (possibly limited) smart meter consumption data. The framework first models a large pool of net load traces by sampling from an appropriately-identified Markov process; this overcomes the drawback of bottom-up demand modeling that fails to properly capture the diversity of customer behavior.
Then, the corresponding battery schedules are computed from a PFA, which was itself trained on solutions to the set of battery scheduling problems of the original customer data. One hundred simulations were carried out per penetration level to capture the uncertainties in the size and location of demand and PV-battery systems. The results show that the PFA reduces the time needed to compute battery schedules by more than \SI{95}{\%}, which makes it feasible to incorporate DER scheduling in a MC framework. The results indicate that uncoordinated PV-battery systems have limited beneficial impact on LV networks, which goes against the conjecture that battery scheduling will serendipitously mitigate the technical problems induced by high PV penetration. Perhaps surprisingly, the inclusion of ToU tariffs in the HEM problem only marginally affects the peak demand compared to SCM, which goes to show that at very high penetration levels, DER scheduling needs to be coordinated by a distribution system operator, using a distribution power flow  \cite{Scott2015,ScottEtal2019,Papavasiliou2018} or peer-to-peer trading with network constraint envelopes \cite{Guerrero2018}.

\section*{References}
\bibliographystyle{elsarticle-num}
\bibliography{bibfile}

\end{document}